\documentclass[a4paper,fleqn]{cas-dc}

\usepackage[numbers]{natbib}
\usepackage{adjustbox}
\usepackage{booktabs}
\usepackage{multirow}
\usepackage{multicol}
\usepackage{subcaption}
\usepackage{url}
\usepackage[colorinlistoftodos]{todonotes}

\newcommand{\probP}{\text{I\kern-0.15em P}}
\newcommand{\expectE}{\text{I\kern-0.12em E}}
\newcommand{\identity}{\text{I\kern-0.12em I}}

\renewcommand{\todo}[2]{#2}
\renewcommand{\textcolor}[2]{#2}

\def\tsc#1{\csdef{#1}{\textsc{\lowercase{#1}}\xspace}}
\tsc{WGM}
\tsc{QE}
\tsc{EP}
\tsc{PMS}
\tsc{BEC}
\tsc{DE}

\begin{document}
\let\WriteBookmarks\relax
\def\floatpagepagefraction{1}
\def\textpagefraction{.001}
\shorttitle{Subject Data Auditing via Source Inference Attack in Cross-Silo Federated Learning}
\shortauthors{Li et~al.}

\title [mode = title]{Subject Data Auditing via Source Inference Attack in Cross-Silo Federated Learning}                      



\author[1]{Jiaxin Li}[
                        orcid=0000-0002-5366-843X,
                        ]
\cormark[1]
\fnmark[1]
\ead{jiaxin.li@studenti.unipd.it}

\credit{Conceptualization, Investigation, Software, Validation, Writing – original draft}

\affiliation[1]{organization={Department of Mathematics, University of Padua},
                addressline={via Trieste, 6}, 
                city={Padova},
                postcode={35131}, 
                country={Italy}}


\author[2]{Marco Arazzi}[%
   orcid=0000-0002-3371-307X,
   ]
\fnmark[1]
\ead{marco.arazzi01@universitadipavia.it}

\credit{Formal analysis, Methodology, Writing – review \& editing}

\affiliation[2]{organization={Department of Electrical, Computer and Biomedical Engineering, University of Pavia},
                addressline={via A. Ferrata, 5}, 
                postcode={27100}, 
                city={Pavia},
                country={Italy}}

\author[2]{Antonino Nocera}[
orcid=0000-0003-2120-2341,
]
\ead{antonino.nocera@unipv.it}

\credit{Formal analysis, Methodology, Supervision, Writing – review \& editing}

\author[1]{Mauro Conti}[
orcid=0000-0002-3612-1934,
]
\ead{conti@math.unipd.it}

\credit{Project administration, Supervision, Writing – review \& editing}


\cortext[cor1]{Corresponding author}
\fntext[fn1]{Equal contribution}


\begin{abstract}
Source Inference Attack (SIA) in Federated Learning (FL) aims to identify which client used a target data point for local model training. It allows the central server to audit clients' data usage. In cross-silo FL, a client (silo) collects data from multiple subjects (e.g., individuals, writers, or devices), posing a risk of subject information leakage. Subject Membership Inference Attack (SMIA) targets this scenario and attempts to infer whether any client utilizes data points from a target subject in cross-silo FL. However, existing results on SMIA are limited and based on strong assumptions on the attack scenario. Therefore, we propose a Subject-Level Source Inference Attack (SLSIA) by removing critical constraints that only one client can use a target data point in SIA and imprecise detection of clients utilizing target subject data in SMIA. The attacker, positioned on the server side, controls a target data source and aims to detect all clients using data points from the target subject. Our strategy leverages a binary attack classifier to predict whether the embeddings returned by a local model on test data from the target subject include unique patterns that indicate a client trains the model with data from that subject. To achieve this, the attacker locally pre-trains models using data derived from the target subject and then leverages them to build a training set for the binary attack classifier. Our SLSIA significantly outperforms previous methods on three datasets. Specifically, SLSIA achieves a maximum average accuracy of 0.88 over 50 target subjects. Analyzing embedding distribution and input feature distance shows that datasets with sparse subjects are more susceptible to our attack. Finally, we propose to defend our SLSIA using item-level and subject-level differential privacy mechanisms. The attack accuracy decreases by 36\% with a utility loss of 20\%, using a subject-level differential privacy budget of 22.




\end{abstract}


\maketitle

\section{Introduction}
Even though Federated Learning (FL)~\cite{mcmahan2017communication, Kairouz_2021_FL} avoids the transition of private data across clients (federation users) to collaboratively train a global model, it is still under the threat of security and privacy attacks. For example, the gradient leakage attack~\cite{Phong_2017_HE_on_FL, Zhu_2019_Deep_leakage, zhao_2020_idlg, Geiping_2020_invert_gradient} recovers training data information or labels of the local models exploiting the shared gradients, the reconstruction attack~\cite{Hitaj_GAN_2017, Wang_2019_User_level} usually trains a GAN to reconstruct the data belonging to a specific label or local client, and, finally, the property inference attack~\cite{zhang2021leakage, Wang_2023_PAPIA} aims to expose the property of the whole training data. The large number of attacks identified by the recent research community indicates that the basic strategy of training local models separately in FL is not enough to protect private data.

With the introduction of regulations like GDPR~\cite{GDPR_citation} and CCPA~\cite{CCPA_citation}, people and governments are paying more attention to auditing privacy data usage. In FL, the Source Inference Attack (SIA)~\cite{SIA_Hu} and Subject Membership Inference Attack (SMIA)~\cite{suri2022subject, Liu_23_SubjectMIA} are two recently proposed attack methods, which can also be considered suitable tools to audit data usage. SIA detects which client trains its local model with a target data point. SMIA checks whether any client (or a specific set of clients) utilizes the data from a target subject for training. However, an honest but curious central server may care more about detecting all the local clients (none, one, or more than one) that train their models with data from the target subject without having access to the specific target dataset used. However, to our best knowledge, there is no such practical and efficient method. Therefore, we propose a novel Subject-Level Source Inference Attack (SLSIA) to achieve this goal.


Our SLSIA distinguishes itself from existing inference attacks, although it shares some similarities with the property inference attack. However, a property inference attack tries to infer the existence or distribution of sensitive property (e.g., voice accent, skin color, income, doctor specialty, product type, and age)~\cite{Ateniese_2015_Hacking, Ganju_2018_PIA, zhang2021leakage, Mahloujifar_2022_PIA_from_poisoning, suri2022formalizing} in the training data of the target model. By contrast, our SLSIA focuses on detecting whether a target local model uses data generated by a subject. The data points belonging to a specific subject are usually not equal to the data points sharing the same property, which distinguishes the property inference and our attack. Besides, our final purpose is to find all the clients using the data from a target subject in a cross-silo FL while holding another set of data points from the subject. In addition, the previous SMIA either can not precisely find which clients train their local models with the data from the target subject~\cite{suri2022subject} or have strong assumptions (such as, they require that considered clients have a local dataset with 10\% data of all subjects, or they interrupt the training of FL, and so forth)~\cite{Liu_23_SubjectMIA}. The DMIA~\cite{Hartmann_2023_DIA} aims to distinguish two datasets that differ only in the contribution of a specific subject, and the remaining data are from the same subjects. However, the local clients holding data from the target subject can still have data from other non-overlapping subjects in cross-silo FL. Hence, different from the DMIA scenario, in the attack scenario we consider, the clients to identify are those having data from the target subject in their local dataset. Still, we cannot assume that the remaining part of the data somehow overlaps. 
As a final remark, we outline that in the scenario considered by the standard instance-level MIAs~\cite{shokri_membership_2017,Gu_2022_CSMIA,Zhang_2023_efficient_MIA}, the member data point is in the training data of one target model. Therefore, the data point is known, and the attack aims to verify whether the target model uses it. Instead, we consider a situation in which some FL clients may use data points (an arbitrary set) from a target source. 
The attacker, located on the server side, does not have access to the specific set of data possibly used by such clients but, instead, has access to other data samples from the same target source.
This complex attack scenario makes our solution novel and more advanced concerning existing related works in this context.

To develop our solution, we analyze recent findings proposed by Song et al.~\cite{Song2020Overlearning}. The authors suggest that the model's intermediate outputs expose the private information of its training data. In particular, the representation of the last layer in the feature extractor overlearns and reveals sensitive attributes of the input data. Following this reasoning, in our solution, we utilize the intermediate embeddings of the local model on a set of data points from the target subject to predict whether the tested model (a local model from one of the clients) has been trained with data from the target subject. The ablation study we carried out and reported in Section ~\ref{different_layer_exploration} shows that the early output of layers is a key factor in developing our attack. 



Our SLSIA pipeline consists of three steps. 
First, the attacker uses data from the target source to pre-train two types of support models: the former exploiting data from the source and the latter exploiting only data from other sources. Once trained, the embeddings produced by these two model types for input data produced by the target source are used to build a binary attack classifier. After that, the binary attack classifier is used to evaluate local models during the first round of FL. The main reason for selecting the first round is that local models trained without the data from the target subject will also learn about the subject as the FL aggregation is executed after each training step, ultimately disturbing the source inference. 
Therefore, during the first round of FL, the central server obtains the embeddings produced by each local model for the same evaluation set of data points from the target subject. Such embedding is then used as input to our binary attack classifier to identify all the clients trained with the data from the target subject.



Our contributions are listed as follows.
\begin{itemize}

\item We propose the SLSIA, a novel source inference attack that can also be used as an efficient and practical method to audit the subject data usage from the perspective of the central server in cross-silo FL. Our SLSIA achieves higher accuracy than previous existing works~\cite{SIA_Hu, suri2022subject} in three datasets.

\item We explore the characteristics of the input data that maximize the performance of our attack. Moreover, through a thorough ablation study, we analyze the main factors allowing our SLSIA attack to be effective. In particular, we demonstrate datasets in which subjects' data have highly different distributions are extremely vulnerable to our SLSIA.

\item We explore item-level and subject-level privacy protection mechanisms against our SLSIA. In general, differential privacy could reduce our attack but also considerably impact the model's utility. The analyzed defenses cannot prevent our SLSIA for most configurations explored in this work.  
\end{itemize}

The remainder of this paper is organized as follows.
Section \ref{sec:background} describes background concepts to introduce our proposal better.
Our proposal, instead, is detailed in Section \ref{sec:Proposal}.
In Section \ref{sec:Exp}, we report all the experiments carried out to validate our proposal.
The obtained results and the analysis of possible defenses against our solution are described in Section \ref{sec:Res}.
Section \ref{sec:RW} analyzes the related literature and identifies the novel aspects of our proposal.
Finally, in Section \ref{sec:Conclusion}, we draw our final conclusion.


\section{Background}
\label{sec:background}
In this section, we introduce the background of FL in Section~\ref{sec:FL}. Then, we recall the SIA in Section~\ref{sec:SIA}. Finally, we review two previous strategies of the SMIA in Section~\ref{sec:SSIA}.

\subsection{Federated Learning}
\label{sec:FL}
In cross-silo FL, a few hundred reliable data silos (organizations or geographically distributed data centers) with powerful computing resources and high-speed connections collectively train a global model without sharing local private data~\cite{Kairouz_2021_FL}. A central server $S$ averages the gradients or weights from local clients (silos) to obtain the global model. Among the averaging strategies, the \textbf{federated averaging} (FedAvg) algorithm~\cite{mcmahan2017communication} is the first and most frequently used algorithm~\cite{QI_2024_FLaberage}. Therefore, we utilize FedAvg to average weights. For each selected local client $C_i$ among $n$ clients in round $j$, its updated weight $W_i^{j}$ is sent to $S$ after its local training with stochastic gradient descent (SGD) on local dataset $D_i$, which is sampled from the whole dataset $D$. The global weight $W^{j}=\sum_{i=1}^{n}I_j(C_i)\frac{|D_i|}{|D|}W_i^{j}$, where $I_j(C_i)=1$ means $C_i$ is selected for updating the global model at round $j$; otherwise, 0. The local client updates the previous global model weight $W^{j-1}$ via multiple local epochs to get the local model weight $W_i^{j}$. For the local epoch, $W_{epoch_k}=W_{epoch_{k-1}}-\eta(\frac{\nabla(\sum_{(x,y)\in D_i}L(f_{W_{epoch_{k-1}}}(x),y))}{\nabla W_{epoch_{k-1}}})$, where $\eta$ is the learning rate, $(x,y)$ is one data point in local dataset $D_i$, $L$ is the loss function (e.g., Cross-Entropy loss for classification task), and $f_{W_{epoch_{k-1}}}(x)$ is the output of model $f$ with parameters $W_{epoch_{k-1}}$ on the input $x$. We set $W_{epoch_0}$ as the previous global model $W^{j-1}$ in the first local epoch. After $m$ local epochs, we set the new local model $W_i^{j}$ from $C_i$ at round $j$ as $W_{epoch_{m-1}}$. In our experiments, we follow the work of Hu et al.~\cite{SIA_Hu} to select all the clients (i.e.,  $I_j(C_i)=1$ for all $C_i$) for updating to simplify the scenario. We leave the scenario that does not select all clients for updating as the future work.


\subsection{Source Inference Attack}
\label{sec:SIA}
For a target data point $z_t=(x_t,y_t)$, the adversary assumes that only local client $C_i$ trains its local model with this target data point. The source inference attack aims to find out this local client. From the strategy of Hu et al.~\cite{SIA_Hu}, they obtain the loss values of $z_t$ on all local models and predict the one with the smallest loss value as the one trained with $z_t$. 
\textcolor{blue}{In particular, they formulate the probability of predicting $z_t$ is in the training data of local client $C_i$ as $\lambda$.}

\begin{equation}
   \probP(s_{ti} = 1) = \lambda.
   \label{eq:probability}
\end{equation}

\textcolor{blue}{Where $s_{ti}$ is the $i$-th element, referred to as the $i$-th client $C_i$, of the one-hot encoding vector $s_t$ (only one element equals 1), which presents the source status of $z_t$.
Given the weights of the $i$-th client $W_i$ and $z_t$, the server rephrases Eq.~\ref{eq:probability} to the posterior probability as follows:}

\begin{equation}
    S(W_i, z_t) = \probP(s_{ti} = 1 | W_i, z_t).
\end{equation}

\textcolor{blue}{The authors derive $S(W_i, z_t)$ from the Bayesian perspective, providing insights on how the
prediction loss of the local clients can be exploited to implement their source inference attack.
The formula of $S(W_i, z_t)$ can be defined as follows:}

\begin{equation}
    S(W_i, z_t) = \expectE_\tau \left[ \sigma \left( log(\frac{\probP(W_i | s_{ti} = 1, z_t, \tau)}{\probP(W_i | s_{ti} = 0, z_t, \tau)}) + \mu_\lambda \right)  \right].
\end{equation}

\begin{equation}
    \tau = \{z_1, \dots, z_{t-1},z_{t+1},\dots, z_{|D|}, s_1, \dots,s_{t-1},s_{t+1},\dots, s_{|D|}\}.
\end{equation}

\textcolor{blue}{Where $\tau$ is the set of the remaining target data points from the whole distribution $D$ and their one-hot encoding vectors containing the source status, $\mu_\lambda = log(\frac{\lambda}{1-\lambda})$, and $\sigma(x) = (1 - e^{-x})^{-1}$ (the sigmoid function). Then, they replace $\probP(W_i | s_{ti} = 1, z_t, \tau)$ with the posterior probability of obtaining the model in the energy-based model~\cite{lecun2006tutorial} to relate $S(W_i, z_t)$ with the loss. By analyzing the optimal source inference from the derived version of $S(W_i, z_t)$, they conclude that the smaller the loss of client $C_i$’s local model on the target data point ($z_t$), the higher the posterior probability that it belongs to client $C_i$.}

\subsection{Subject Membership Inference Attack}
\label{sec:SSIA}
In the conventional membership inference attack, the adversary aims to infer whether a specific data point is in the training data of the target model~\cite{shokri_membership_2017}. Considering local datasets are generated by multiple subjects (e.g., various devices or individuals) in cross-silo FL, the adversary wants to know whether any local dataset $D_i$ contains data from a target subject $s_t$, which is formulated as the SMIA in the work of Suri et al.~\cite{suri2022subject}.
\textcolor{blue}{For $n$ clients in FL to update the global model weight $W^j$ for $r$ rounds, where $j$ denotes the $j$-th round, the authors propose \textbf{Loss-Threshold} and \textbf{Loss-Across-Rounds} attacks. In the \textbf{Loss-Threshold} attack, they record the loss values ($L$) for each data point $(x,y)$ sampled from the data of the target subject ($D_{s_t}$) and check if the loss values are lower than a given threshold $\lambda$ as follows:}

\begin{equation}
    c = \sum_{(x,y) \in D_{s_t}} \identity[L(f_{W^j}(x),y) \leq \lambda].
\end{equation}

\textcolor{blue}{Where $\identity$ is the identity function. 
With the obtained $c$, the attacker predicts the data of the target subject ($s_t$) is used in FL if $c$ is non-zero or the attacker defines an additional threshold for $c$ based on the metric he wants to maximize.
In the \textbf{Loss-Across-Rounds} attack, the attacker sums the loss values of data points from the target subject as $c_j = \sum_{(x,y) \in D_{s_t}} L(f_{W^j}(x),y)$ in round $j$. Then, the attacker records the number of times the summed loss decreases concerning the previous round.}

\begin{equation}
    c = \sum_{j=1}^r \identity[c_j < c_{j-1}].
    \label{c_eq}
\end{equation}

\textcolor{blue}{The attacker can now derive a threshold for $c$ in Eq.~\ref{c_eq} to determine whether the data from the target subject ($s_t$) is used in FL.} In the work of Liu et al.~\cite{Liu_23_SubjectMIA}, instead, they care about whether a set of concerned clients rather than any clients contain the data from the target subject in FL with strong assumptions (10\% data of all subjects in concerned clients, interrupt the training of FL, and datasets without a concept of the subject). From previous definitions~\cite{suri2022subject, Liu_23_SubjectMIA, Hartmann_2023_DIA}, the SMIA differentiates between models trained on datasets sampled from two distributions, the only difference of which is that one has the data from the target subject while the other does not.


\section{Our Subject-Level Source Inference Attack}
\label{sec:Proposal}

Our SLSIA targets detecting all the local clients that train their local models with the data from the target subject from the perspective of an honest but curious central server. We explain the attack objective of our SLSIA in Section~\ref{sec:attack_objective}. Then, we formulate the threat model in Section~\ref{sec:threat_model}. Finally, we detail the methodology of applying our SLSIA in Section~\ref{sub:methodology}.

\subsection{Attack Objective}
\label{sec:attack_objective}
In cross-silo FL, an honest but curious central server $S$ updates global model weight $W^j$ at each round $j$ with local model weights $W_i^j(i=1,...,n)$ from local clients ($n=10$ in default). Each local dataset $D_i$ is sampled from the whole distribution $D$, which consists of data from $k$ subjects. In FL, the data from a subject might only be available to a few local clients as a requirement for privacy projection (e.g., the patient's health records are only available in the post-diagnostic hospitals and are usually not allowed to be shared). In the worst-case scenario, two local datasets typically do not sample from the same subjects and obtain data from entirely different subjects. To regulate and audit the data usage of local clients, an honest but curious central server $S$ wants to figure out all the local clients that utilize the data from a target subject ($s_t$) to train their local models. In other words, we distinguish two local models, one containing the data from the target subject for training while the other does not. Besides, the remaining data from those two local datasets might be sampled from different subjects, which means the data from the target subject is not the only difference. It is one aspect that distinguishes our attack from the subject membership and property inference attacks.

\subsection{Threat Model}
\label{sec:threat_model}
To correctly detect all the local clients trained with the data from the target subject as an honest but curious server, we assume that the server knows (1) the structure of the local model, (2) hyper-parameters of local training, (3) reserved data points from the target subject (not used in local training), (4) data points from other subjects which are not used in local training, and (5) weights of local models (white-box). As the central server, obtaining the structures and weights of local models is reasonable as the server needs to update the global model. The server can determine the hyper-parameters for local clients before FL or apply the stealing attack to get hyper-parameters~\cite {Wang_2018_hypersteal}. The data points from the target subject are necessary for the attack to be applied. \textcolor{blue}{The central server holds a set of data points ($D_{s_t}^{s}$) from the target subject set ($D_{s_t}$) that does not overlap with the training data ($D_{s_t}^{c}$) used by $m$ target local clients (${C_t=\{C_{t_1},\dots, C_{t_m}\}}$) it wants to detect. Besides, the server collects a set of datasets ($\{D_{r_1,\dots,r_r}\}$) from other random subjects (${\{r_1,\dots,r_r\}}$).
The server exploits the mentioned knowledge to design a strategy for detecting whether a client uses the data from the subject target ($s_t$).
Our approach to implementing SLMIA is presented in the following Section~\ref{sub:methodology}}.




\subsection{Methodology}
\label{sub:methodology}


\textcolor{blue}{
As we present in Section~\ref{sec:background}, SIA predicts the only target client as the one with the smallest loss on the target data point. SMIA compares the losses of data points from the target subject with a threshold or accumulates those losses among the training epochs to observe the number of decreases of the summed loss. We can not directly apply previous approaches to our scenario, in which there can be more than one or no target client, the total number of them is unknown to the server, and the final purpose is to detect all target clients. We explain two transferred methods from previous works in Section~\ref{sub:attackPerfromance} for comparison. The target model means the local model trained with one target data point (SIA), the global model trained with the data from a target subject (SMIA), or the local model trained with the data from a target subject (our SLSIA).}


\textcolor{blue}{Our SLSIA is based on the data from the target subject ($s_t$). As we mentioned in Section~\ref{sec:threat_model}, we divide the data point set ($D_{s_t}$) from the target subject into two non-overlapping sets, $D_{s_t}^{s}$ (for pre-training and evaluation) and $D_{s_t}^{c}$ (for FL). Then, we further split $D_{s_t}^{s}$ into two sets, $D_{s_t}^{p}$(for pre-training) and $D_{s_t}^{e}$ (for evaluation). We explain the default proportions of $D_{s_t}^{p}$, $D_{s_t}^{e}$, and $D_{s_t}^{c}$ over $D_{s_t}$ in Section~\ref{sec:models_and_settings}.}

\textcolor{blue}{In our proposed attack, we employ a binary classifier as Attack Model $A_M$ to detect whether the embeddings of the data of target subject $s_t$ are obtained from a local model (with the weight $W_i^1$ from the local client $C_i$) has been trained with the data of $s_t$ in the first round. To train our Attack Model $A_M$ offline before the FL process starts, the server pre-trains $N^{pre}$ pre-trained models $M^{pre}$, with the same architecture as the global model, to generate the embeddings $H^{pre}$ as the input features of $A_M$. The $H^{pre}$ comprises two sets of embeddings, one generated by $\frac{N^{pre}}{2}$ pre-trained models partially trained with the data from $s_t$ ($D_{s_t}^{p}$) and one obtained from other $\frac{N^{pre}}{2}$ pre-trained models trained on data points only from random subjects.}

\textcolor{blue}{In detail, the pre-training of models, $M^{pre}$, is performed by simulating $N^{pre}$ different datasets $D^{pre}$ using the data from the target subject ($D_{s_t}^{p}$) and the data from random subjects ($\{D_{r_1,\dots,r_r}\}$). Among them, $D_{ti}^{pre}$ contains the data from the target subject ($D_{s_t}^{p}$) and a similar number of data points from a random subject. $D_{ri}^{pre}$ includes a similarly equal number of  data points from two random subjects separately. It means the number of data points in $D_{ti}^{pre}$ and $D_{ri}^{pre}$ is similar.
\begin{equation}
    D^{pre} = \{D_{t1}^{pre}, \dots,  D_{t\frac{N^{pre}}{2}}^{pre}, D_{r1}^{pre}, \dots,  D_{r\frac{N^{pre}}{2}}^{pre}\}.
\end{equation}
After training with $D^{pre}$, the server trains $N^{pre}$ pre-trained models $M^{pre}$ to extract the data of training Attack Model $A_M$.
\begin{equation}
    M^{pre} = \{M_{t1}^{pre}, \dots,  M_{t\frac{N^{pre}}{2}}^{pre}, M_{r1}^{pre}, \dots,  M_{r\frac{N^{pre}}{2}}^{pre}\}.
\end{equation}
$M_{ti}^{pre}$ is the model trained on $D_{ti}^{pre}$, and $M_{ri}^{pre}$ is trained on $D_{ri}^{pre}$. We call $M_{ti}^{pre}$ the \textbf{target pre-trained model} and $M_{ri}^{pre}$ the \textbf{random pre-trained model} separately. With $N^{pre}$ pre-trained models $M^{pre}$, the server generates the embeddings $H^{pre}$ via querying the pre-trained models with an evaluation set of data points from the target subject ($D_{s_t}^{e}$). If the embedding is obtained from the target pre-trained model ($M_{ti}^{pre}$), the server assigns the label $y^t=1$ ("in") to this embedding. If the embedding is generated by the random pre-trained model $M_{ri}^{pre}$, the label is $y^r=0$ ("out").}
\begin{gather}
    H_{ti}^{pre} = M_{ti}^{pre}(D_{s_t}^{e}); \ H_{ri}^{pre} = M_{ri}^{pre}(D_{s_t}^{e}). \label{eq_1}\\
    H^{pre} = \{H_{ti}^{pre}, H_{ri}^{pre}\} \ where \ i \in [1,\dots, \frac{N^{pre}}{2}]. \label{eq_2}
\end{gather}
\textcolor{blue}{With the embeddings and labels, the server trains $A_M$ via supervised learning in Eq.~\ref{eq_AM}. After being trained to perform well, $A_M$ precisely outputs 1 and 0 for embedding separately from the  target and random pre-trained models.}
\begin{equation}
    A_M = \arg \min_{M} \left( \sum_1^{\frac{N^{pre}}{2}} L(M(H_{ti}^{pre}), y^t) \ + \ \sum_1^{\frac{N^{pre}}{2}} L(M(H_{ri}^{pre}), y^r) \right).\label{eq_AM}
\end{equation}
\textcolor{blue}{$L$ is the loss function used to train the attack model $A_M$.  With $A_M$, the server generates the embeddings $H_i^{eval} = \{h_{i1}^{eval}, \dots, h_{i|D_{s_t}^{e}|}^{eval}\}$ with the same evaluate set of data points from the target subject ($D_{s_t}^{e}$) via querying the local model ($W_i^1$) from the local client $C_i$ in the first round. }
\begin{equation}
H_i^{eval} = Embed_{f_{W_i^1}}(D_{s_t}^{e}).
\end{equation}
\textcolor{blue}{$Embed$ is the intermediate embedding output of the local model. Then, the server feeds generated embeddings $H_i^{eval}$ into the attack model $A_M$. If more than $50\%$ of the embeddings $H_i^{eval}$ are classified as "in" ($1$) by Attack Model $A_M$, the server predicts that local client $C_i$ trains its local model with the data from the data of the target subject ($D_{s_t}^{c}$).}
\begin{gather}
    T_{C_i} = \sum_{x=1}^{|D_{s_t}^{e}|} A_M(h_{ix}^{eval}). \\ 
    C_i \in C_t \text{ if } T_{C_i} \geq \frac{|H_i^{eval}|}{2}.
\end{gather}
\textcolor{blue}{$T_{C_i}$ denotes the target score to evaluate if $C_i$ is one target local client, i.e., a part of target local clients $C_t$. After evaluating local models from all clients, the adversary can find all the target local models trained with the data from the target subject.}

\begin{figure}[ht!]
  \centering
  \includegraphics[scale=0.255]{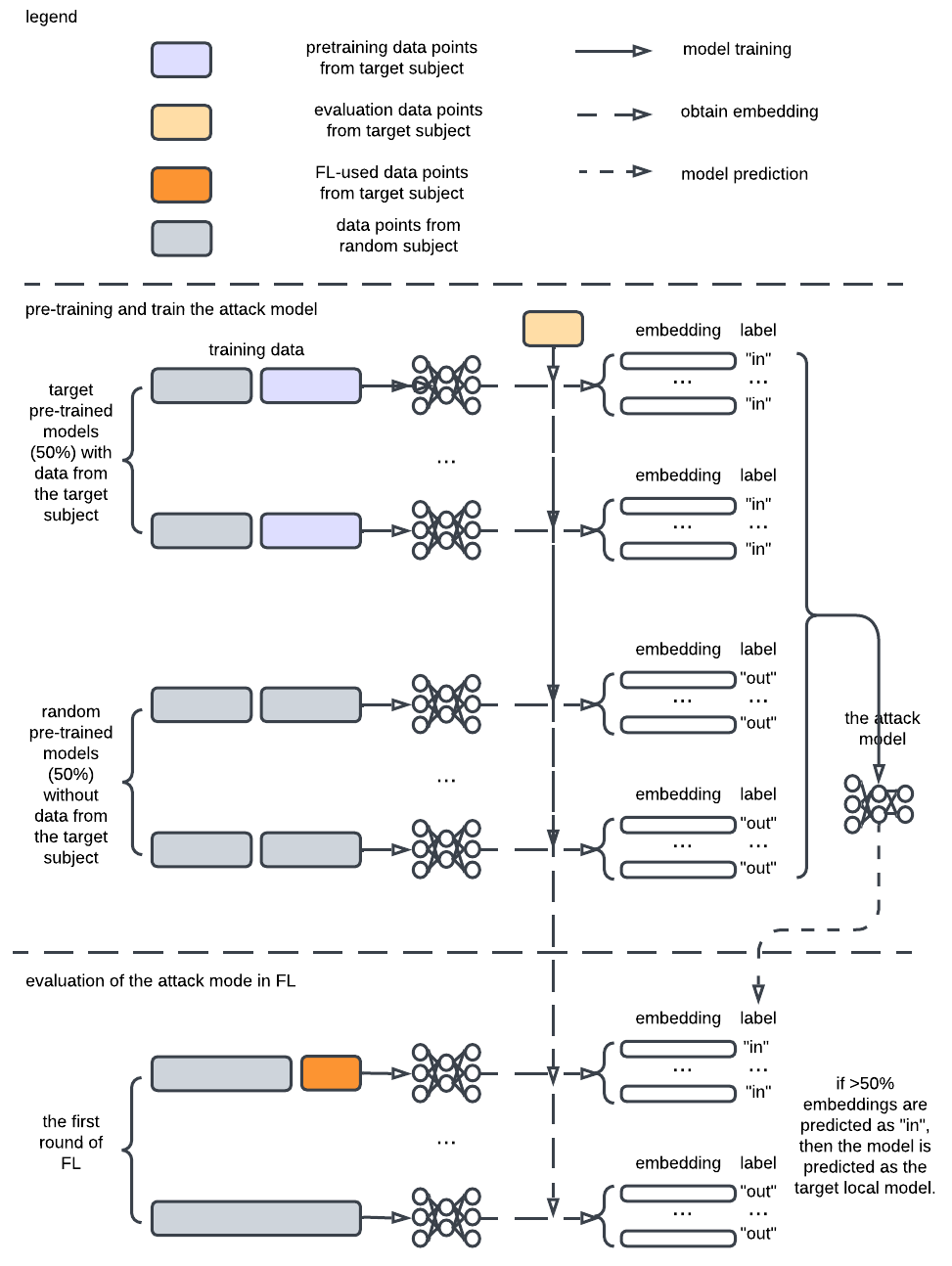}
  \caption{The pipeline of our subject-level source inference attack includes three stages: pre-training models, extracting embeddings to train the attack model, and evaluating the local models.}
  \label{our_SLSIA_pipeline}
\end{figure}

The general pipeline of our SLSIA is exposed in Figure~\ref{our_SLSIA_pipeline}. From the picture, we can observe the three stages mentioned before, including pre-training models, extracting embeddings to train the attack model, and evaluating the local models to predict which are trained with the data from the target subject. Besides, the pre-trained and local FL models are initialized from the same global model weight, $W^0$, to avoid the impact of model initialization on the learning of the pre-trained models.

\section{Experiments}
\label{sec:Exp}
To verify the effectiveness of our proposed SLSIA, we experiment on three datasets and three types of models. We describe the datasets used in experiments in Section~\ref{sec:datasets}. For models, settings to train models, and partial default settings of our SLSIA, we expose them in Section~\ref{sec:models_and_settings}. Finally, we detail the evaluation metrics for measuring the attack performance in Section~\ref{sub:evaluation_metrics}.

\subsection{Datasets}
\label{sec:datasets}
Following the work of Suri et al.~\cite{suri2022subject}, we select FEMNIST and Shakespeare datasets with a clear concept of the subject for the attack. The FEMNIST is a federated extended MNIST~\cite{Deng_2012_MNIST} \textcolor{blue}{that not only contains classes characterized by handwritten digits and letters but also includes information on writers who contributed to the collection of data points. This additional writer information makes the dataset perfectly fit our scenario, in which we consider each writer a subject.} Here, we only use the digits following the MNIST, which only contains the digits. Besides, we can reduce the impact of the internal difference of the dataset caused by combining the digits and letters to the classification and the attack.

\textcolor{blue}{As the second dataset, we selected the Shakespeare dataset from the LEAF~\cite{caldas2018leaf} family, a repository that provides different federated benchmark datasets.
The Shakespeare dataset contains dialogues of characters in the classic William Shakespeare plays, and the characters can be considered different subjects.
In this case, the main task of FL is the next-word prediction. 
We opted for this task due to the limitations of alternatives applicable to this dataset in our subject-centric scenario, like the next-character task, as highlighted by other works~\cite{suri2022subject}.} 
Besides, we excluded characters with less than 400 words to guarantee the data points for each character and eliminate the impact of data inadequacy while being selected as the target subject of the attack. Each data point in the Shakespeare dataset is a continuous sequence of 32 words with a label of the next word. Finally, we obtain data on 404 characters in the Shakespeare dataset. 

Apart from those two already existing datasets, we also construct a Synthetic dataset following the guide of~\cite{suri2022subject}.  We model each subject as a random (and valid) mean and covariance matrix in a multivariate Gaussian distribution \textcolor{blue}{to obtain well-separated and characterized subjects. } In particular, we enforce a minimum pair-wise (L2 $>$ 0.35) separation between all subject distributions’ means to avoid overlap. The label of a generated data point \textcolor{blue}{is the outcome of an XOR operation between a list of indicator values, each representing whether the specific feature is larger than 0. For the input feature $x=[x_1,\dots,x_p]$, the label $y=\oplus_{i=1}^{p}\identity(x_i\geq0)$, where $\oplus$ is the XOR operation and $\identity$ is the identity function.}\todo{Can we add a little more details on this?}
We construct a Synthetic dataset with 200 subjects, each with 400 data points. The length of the feature is 60, and the number of categories is two due to the range of XOR values. We summarize the statistics of each dataset in Table~\ref{dataset_infor}.

\begin{table}[ht]
\centering
\caption{Statistical information of datasets.}
\label{dataset_infor}
\begin{adjustbox}{max width=0.34\textwidth}
\begin{tabular}{ccc}
\toprule
Dataset  & Feature Length  & Subject Number \\
\cmidrule{1-3}
FEMNIST&  1*28*28  & 3,580 \\
Shakespeare & 32 & 404\\
Synthetic & 60       & 200  \\
\bottomrule
\end{tabular}
\end{adjustbox}
\end{table}


\subsection{Models and Settings}
\label{sec:models_and_settings}
\textcolor{blue}{This section is devoted to presenting the architectures of the models and the settings used to conduct our experimental campaign. For each dataset, we defined a specific architecture shared by the global, local, and pre-trained models used in our attack, specific for the domain that characterizes their features.}

In particular, for FEMNIST, we use a convolutional neural network (CNN), as the one in~\cite{SIA_Hu}, \textcolor{blue}{characterized by two layers with kernel size equal to $5$ and, respectively, $32$ and $64$ filters. A max-pooling layer of kernel 2 follows each of the two layers. The classifier on top is a multilayer perceptron (MLP) network composed of three layers, of which the first two are characterized by $512$ and $128$ neuron units and ReLu activation function, and the final one exits with a softmax function. }
\textcolor{blue}{For Shakespeare, we considered an embedding network characterized by a one-layer long short-term memory (LSTM~\cite{pytorch_lstm_2024}) followed by a classification layer of $64$ neuron units as our next word prediction model.}
\textcolor{blue}{The size of the embedding generated in output by the LSTM equals $64$, which is also the size of the internal hidden states.}
We take the last hidden state of LSTM as the input of the classification layer. 
\textcolor{blue}{For Synthetic, following the work of ~\cite{SIA_Hu}, we employ an MLP network with one hidden layer with 200 units and a ReLu activation function.}

We call the above three models CNN, LSTM, and MLP for convenience, except for additional illustration. We utilize SGD as an optimizer with a learning rate of 0.01 and a momentum of 0.9 for all experiments. 
\textcolor{blue}{For the training of the local model on the client side, we set the batch size equal to $12$, and at each round, the client performs five local epochs before sending the updates to the server.}
\textcolor{blue}{In the same way, we pre-train $N^{pre}$ models used by the server to generate the embeddings useful for the training of the attack model.}

\textcolor{blue}{We select two types of models for the binary attack model: 1-dimensional CNN and support vector machine (SVM). The CNN-based attack model comprises two 1-dimensional convolution layers with $4$ and $8$ filters and a classification layer with Softmax function. Each 1-dimensional convolution layer is followed by a 1-dimensional max pool with a kernel size of three and a Batch Normalization layer. The SVM classifier utilizes the default settings in the \textbf{scikit-learn}~\cite{sklearn_svc_2024} library.} We optimize the CNN-based attack model via Adam with a learning rate of 0.0001 and a weight decay of 0.1. The batch size is 16, and the training epoch is 100. As the loss function, we consider the coss-entropy loss. 

\textcolor{blue}{Finally, we explain the partial default settings of applying our attack strategy here. The number of local clients ($n$) is 10, the same as the work of Hu et al.~\cite{SIA_Hu}. The number of pre-trained models ($N^{pre}$ ) is 20. The reason for pre-training 20 models is to guarantee the embedding dataset size for training the attack model. The number of target local clients ($m$) trained with the data from the target subject is 5, half of the total local clients, as we explain in Section~\ref{sub:evaluation_metrics}. To divide the data points from the target subject, we split 25\% of $D_{s_t}$ as $D_{s_t}^c$ used by $m$ target local clients, which means $D_{s_t}^s=75\%D_{s_t}$. As the central server knows the target subject, he can collect data points from the target subject by himself. Hence, it is reasonable that the server holds most (75\%) data points of the target subject. $D_{s_t}^s$ is split into $D_{s_t}^p=50\%D_{s_t}$ (for pre-training) and $D_{s_t}^e=25\%D_{s_t}$ (for extracting the embeddings). As the number of pre-trained models is twice that of the local clients of FL in our setting, we utilize more data from the target subject for pre-training. Besides, we illustrate the default settings and have an ablation study for the percentage of the data from the target subject in the target local client, the embedding layer, and the local training epoch in Section~\ref{sub:ablation_study}.}

\subsection{Evaluation metrics}
\label{sub:evaluation_metrics}

\textcolor{blue}{In this section, we discuss the evaluation metrics we defined to assess the performance of our attack.}
In the work of Hu et al.~\cite{SIA_Hu}, they utilize the Attack Success Rate (ASR) to measure the performance of the SIA on an evaluation group of target data points. They define the ASR as the fraction of the target data points whose source status is correctly identified by the server in the evaluation group. Their ASR is similar to the recall, which is the percentage of correctly predicted positive instances among all the actual positive instances. To measure the performance of our SLSIA, we define metrics from the perspective of each target subject. Assuming the data from the target subject $s_t$ is in the training data of partial of $n$ clients, we have a ground-truth source list, $[0,\dots,1]$, comprised of $n$ indicator values; each value is 0 or 1. If the $i$-th indicator value is 1, the local client $C_i$ trains its model with the data from the target subject $s_t$; otherwise, it does not.

After applying our SLSIA, we obtain the potential clients \textcolor{blue}{that train their local model using the data from the target subject}, which is also formulated as a potential source list, $[0,\dots,1]$, having the same length as the ground-truth source list and representing the same meaning with the value at the same position. 
\textcolor{blue}{Comparing the ground-truth source list with the potential source list predicted by our attack, we calculate each target subject's accuracy, precision, recall, and F1.
Then, distinct metrics of target subjects are averaged to produce the performance of our attack, and we compare it with previous methods.}
\textcolor{blue}{When we compare the ground-truth source list and the potential source list by calculating the accuracy in the classification task, the number of "1" in the ground-truth source list determines whether the calculated accuracy is balanced. To obtain a balanced accuracy value and guarantee the fairness of evaluation, we set the number of local clients trained with the data from the target subject as half of the total number of clients as default without additional illustration, which means that half of the ground-truth source list is "1".} \todo{This sentence is not clear to me} Because we implement our SLSIA at the first round of FL, setting half of the total number of clients to be trained with the data of the target subject will not impact the attack performance as the FL aggregation is applied after the first FL round.


\section{Results and Discussions}
\label{sec:Res}

\textcolor{blue}{This section presents and analyzes the results of our SLSIA. In the first place, we report the performance of the local model on the original task after the first round of the federated process in Section~\ref{sub:originalPerfromance}. Then, in Section~\ref{sub:attackPerfromance}, we present the main results of our attack using the evaluation metrics presented in Section~\ref{sub:evaluation_metrics}. Section~\ref{sub:subject_analysis} investigates the differences between the embeddings containing information about the target or random subjects. In Section~\ref{sub:ablation_study}, we report an ablation study related to the percentage of the data from the target subject in target local models, the embedding layer, and the number of local training epochs. Finally, in Section~\ref{sub:defense}, we test our strategy against item-level and subject-level differential privacy.}

\subsection{Performance of Original Task}
\label{sub:originalPerfromance}
As we implement our SLSIA at the first round of FL, the original task's performance is inevitably relatively low (even close to random guess) compared to the well-trained models. For FEMNIST, the accuracy of the local clients on the digit classification task is about 6\% to 18\% (total of 10 digits), relatively higher than random guess (10\%). For Shakespeare, the next word prediction accuracy of local models in the first round is about 1\% to 4\%, which is more significant than the random guess due to the size of the word dictionary being 30353 in our filtered Shakespeare dataset. For Synthetic, the XOR value prediction with the artificially generated features has an accuracy of about 40\% to 80\%, higher than the random guess (50\%). As the mentioned accuracy is obtained in the first round of FL, it is expected to have a relatively lower performance, corresponding to our observations.  

\subsection{Attack Performance}
\label{sub:attackPerfromance}
We compare our SLSIA with two methods inspired by previous works to fit our purpose of detecting all the clients trained with the data from the target subject. We name them (1) \textbf{Avg Loss} and (2) \textbf{Min Loss Time} methods. The \textbf{Avg Loss} method compares the average loss values of the evaluation data points from the target subject ($D_{s_t}^e$) on all the local models. 
\textcolor{blue}{Following the strategy presented by Hu et al.~\cite{SIA_Hu}, this approach considers potential target local clients as the ones that, using their local model updates, produce the smallest average loss across the clients.}
Instead, the \textbf{Min Loss Time} method counts the times each local client obtains the minimum loss value among the evaluation data points ($D_{s_t}^e$). Then, the local clients who get the minimum loss value more frequently are determined as potential target local clients. This is motivated by setting a threshold on the loss value of the data from the target subject in the previous work~\cite{suri2022subject} . 
\textcolor{blue}{The method described in the original paper relies on the intuition that if the local model has been trained using data points from the target subject, the resulting loss on the data of the target subject will be lower than a given threshold concerning random data points from other subjects.}
\textcolor{blue}{In our implementation, instead of setting a fixed threshold, we count how many times a client registers the lowest loss on the evaluation data points of the target subject ($D_{s_t}^e$) after one training round of FL.}
To make a fair comparison, we only implement those two methods in the first round as our SLSIA. 
\textcolor{blue}{In our scenario, we allow the possibility that more than one client can train their local clients using samples from the subject target. Suppose we follow the work of Hu et al.~\cite{SIA_Hu} to determine the one with the lowest average loss or maximum number of times to obtain the minimum loss in the \textbf{Avg Loss} and \textbf{Min Loss Time} methods as the target local client. In that case, we can not find all the target local clients, which is unsuitable for our scenario. Besides, finding a threshold suitable to determine which clients are target local clients for all training tasks is not easy. Hence, to make \textbf{Avg Loss} and \textbf{Min Loss Time} methods applicable to our scenario, we assume that the attacker knows precisely the number of target local clients, which is a strong assumption in a practical scenario. Our solution does not require this prerequisite, allowing us to relax this assumption and making it more realistic compared to the state of the art.}\todo{Please check if this last sentence is true}

\begin{figure*}
    \centering
    \begin{subfigure}[b]{0.24\textwidth}
        \centering
        \includegraphics[scale=0.42]{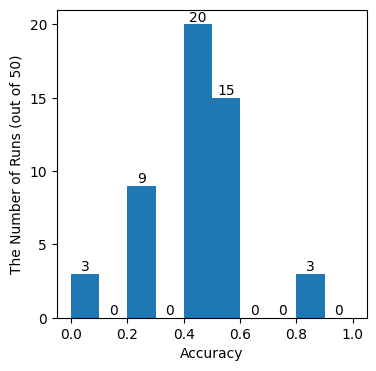}
        \caption{Avg Loss-FEMNIST}
        \label{Avg Loss-FEMNIST}
    \end{subfigure}
    \begin{subfigure}[b]{0.24\textwidth}
        \centering
        \includegraphics[scale=0.42]{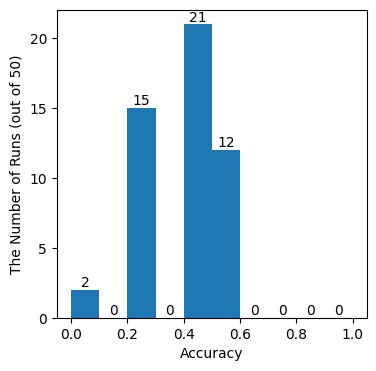}
        \caption{Min Loss Time-FEMNIST}
        \label{Min Loss Time-FEMNIST}
    \end{subfigure}
    \begin{subfigure}[b]{0.24\textwidth}
        \centering
        \includegraphics[scale=0.42]{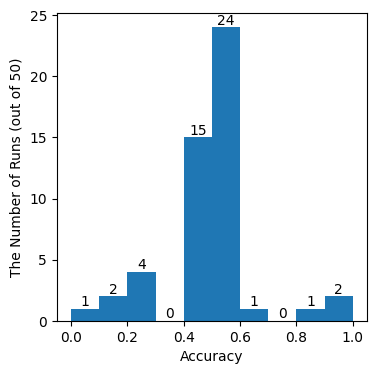}
        \caption{SLSIA (CNN)-FEMNIST}
        \label{SLSIA (CNN)-FEMNIST}
    \end{subfigure}
    \begin{subfigure}[b]{0.24\textwidth}
        \centering
        \includegraphics[scale=0.42]{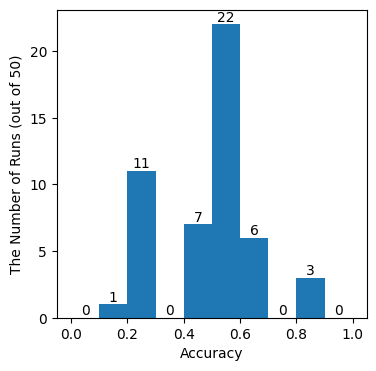}
        \caption{SLSIA (SVM)-FEMNIST}
        \label{SLSIA (SVM)-FEMNIST}
    \end{subfigure}
    \begin{subfigure}[b]{0.24\textwidth}
        \centering
        \includegraphics[scale=0.42]{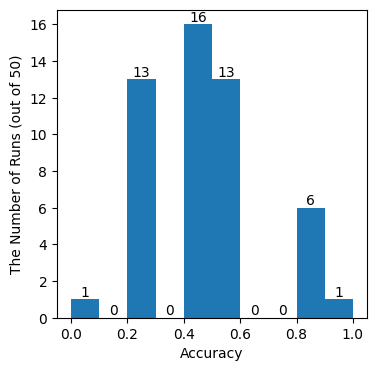}
        \caption{Avg Loss-Shakespeare}
        \label{Avg Loss-Shakespeare}
    \end{subfigure}
    \begin{subfigure}[b]{0.24\textwidth}
        \centering
        \includegraphics[scale=0.42]{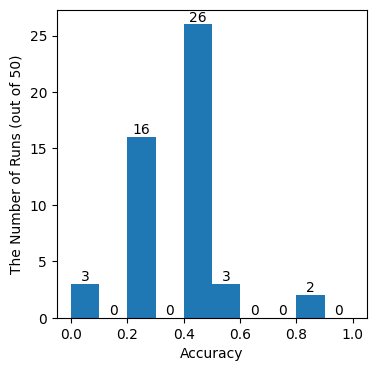}
        \caption{Min Loss Time-Shakespeare}
        \label{Min Loss Time-Shakespeare}
    \end{subfigure}
    \begin{subfigure}[b]{0.24\textwidth}
        \centering
        \includegraphics[scale=0.42]{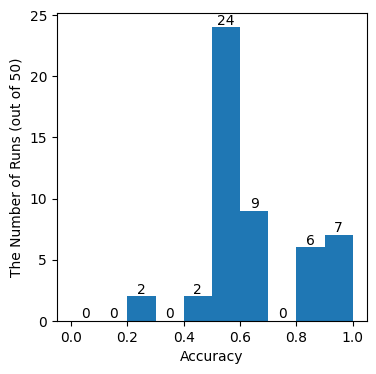}
        \caption{SLSIA (CNN)-Shakespeare}
        \label{SLSIA (CNN)-Shakespeare}
    \end{subfigure}
    \begin{subfigure}[b]{0.24\textwidth}
        \centering
        \includegraphics[scale=0.42]{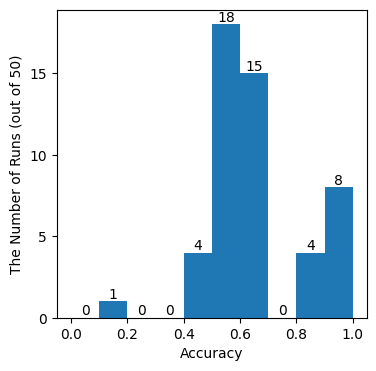}
        \caption{SLSIA (SVM)-Shakespeare}
        \label{SLSIA (SVM)-Shakespeare}
    \end{subfigure}
    \begin{subfigure}[b]{0.24\textwidth}
        \centering
        \includegraphics[scale=0.42]{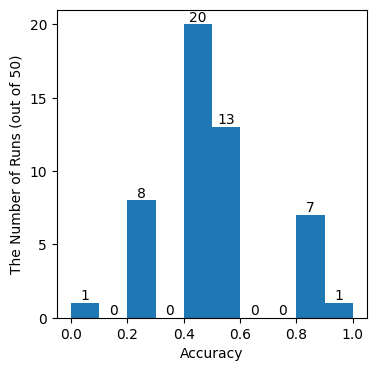}
        \caption{Avg Loss-Synthetic}
        \label{Avg Loss-Synthetic}
    \end{subfigure}
    \begin{subfigure}[b]{0.24\textwidth}
        \centering
        \includegraphics[scale=0.42]{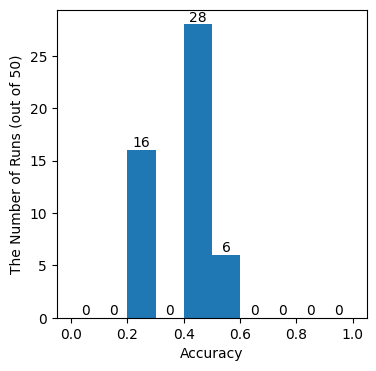}
        \caption{Min Loss Time-Synthetic}
        \label{Min Loss Time-Synthetic}
    \end{subfigure}
    \begin{subfigure}[b]{0.24\textwidth}
        \centering
        \includegraphics[scale=0.42]{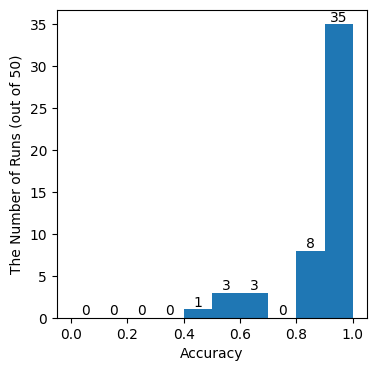}
        \caption{SLSIA (CNN)-Synthetic}
        \label{SLSIA (CNN)-Synthetic}
    \end{subfigure}
    \begin{subfigure}[b]{0.24\textwidth}
        \centering
        \includegraphics[scale=0.42]{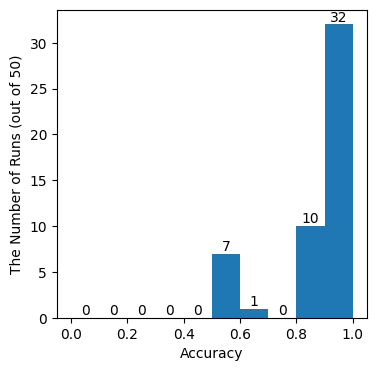}
        \caption{SLSIA (SVM)-Synthetic}
        \label{SLSIA (SVM)-Synthetic}
    \end{subfigure}
    \caption{Attack accuracy distribution of 50 runs (subjects). The caption of each sub-figure is "attack method name-dataset." In the sub-figure, the x-axis is the accuracy range of one run, and the y-axis is the number of runs obtaining an attack accuracy within the accuracy range.}
    \label{attack_accuracy_distribution_50subjects}
\end{figure*}

To compare the performance of the attacks, we randomly select 50 subjects as the target subjects of 50 runs in each dataset and obtain the accuracy, precision, recall, and F1 of each subject under each method as mentioned in Section~\ref{sub:evaluation_metrics}. Figure~\ref{attack_accuracy_distribution_50subjects} shows the attack accuracy distribution while attacking 50 subjects with various approaches. 
Our two methods (CNN-based and SVM-based SLSIA) obtain higher accuracy than others. In particular, the number of runs (subjects) with an accuracy higher than 0.9 reaches 35 and 32 with our methods under the Synthetic dataset, which is significantly higher than the other two methods (\textbf{Avg Loss}: 1 and \textbf{Min Loss Time}: 0). For the Shakespeare dataset, \textbf{SLSIA (CNN)} has seven runs with an accuracy higher than 0.9 while \textbf{SLSIA (SVM)} has eight runs, compared to 1 for the \textbf{Avg Loss} and 0 for the \textbf{Min Loss Time} methods. In the FEMNIST dataset, only \textbf{SLSIA (CNN)} has two runs with an accuracy higher than 0.9. It shows that the performance of our SLSIA is better than that of the other two methods. Besides, the attack performance is related to the datasets themselves from the results of the three datasets. \textcolor{blue}{For the FEMNIST dataset, this kind of performance is expected due to the high similarity between the styles of the different writers. In Section~\ref{sub:subject_analysis}, we better prove this point by providing an in-depth analysis of the subject property in each dataset concerning the resulting attack performance.}

Besides the accuracy, we also observe the higher values of the other three metrics in our methods compared with the other two approaches. Table~\ref{avg_metric_value_over_50_subejcts} shows the average value of each metric list obtained from 50 runs (subjects) under each method. \textcolor{blue}{The average metric values (accuracy, precision, recall, and F1) of both \textbf{Avg Loss} and \textbf{Min Loss Time} are respectively about 0.45 and 0.35, which is close to random guess.}
The inefficiency of those two methods indicates that the model trained with the data from the target subject will not obtain a lower loss and will not frequently get a minor loss among all local models on evaluation data points from the target subject. There are two reasons behind this phenomenon. First, the evaluation data points from the target subject are not included in the weight update during FL training, which can not guarantee a lower loss and frequent minor loss. The second reason is that the local model trains one round of FL before attacking, which makes the local model learn insufficient information from the target subject's data ($D_{s_t}^c$). It is hard to determine the FL round where the target local clients learn better from the target subject data than random local clients. It is also one of the reasons that led us to perform our SLSIA at the first FL round.

The average metric values of our SLSIA are mostly higher than the \textbf{Avg Loss} and \textbf{Min Loss Time} methods. For example, under the Synthetic dataset, \textbf{SLSIA (CNN)} obtains an average accuracy of 0.88, compared to 0.45 and 0.38 of the two previous methods. The average precision, recall, and F1 of \textbf{SLSIA (CNN)} are 0.91, 0.87, and 0.87, significantly higher than previous methods. The attack performance of our SLSIA on Shakespeare and Synthetic datasets empirically proves the effectiveness of our attack. For the FEMNIST dataset, our SLSIA has a higher average accuracy but is lower on the other three metrics, especially in F1, compared with the two previous methods. We attribute the low performance of our SLSIA on the FEMNIST to the fact that most subjects are not distinguishable enough, which is also mentioned in Section~\ref{sub:subject_analysis}.

\begin{table*}[ht]
\centering
\caption{The averaged metric values of 50 runs (subjects) under various methods.}
\label{avg_metric_value_over_50_subejcts}
\begin{adjustbox}{max width=0.98\textwidth}
\begin{tabular}{ccccccccccccccccc}
\hline
\addlinespace 
\multirow{3}{*}{Dataset}  & \multicolumn{4}{c}{Averaged Accuracy} & \multicolumn{4}{c}{Averaged Precision} & \multicolumn{4}{c}{Averaged Recall} & \multicolumn{4}{c}{Averaged F1} \\ 
\cmidrule(lr){2-5} \cmidrule(lr){6-9} \cmidrule(lr){10-13} \cmidrule(lr){14-17}
 &  \multirow{2}{*}{\begin{tabular}[c]{@{}c@{}}Avg Loss\end{tabular}} &\multirow{2}{*}{\begin{tabular}[c]{@{}c@{}}Min Loss\\ Time\end{tabular}} &\multirow{2}{*}{\begin{tabular}[c]{@{}c@{}}SLSIA\\ (CNN)\end{tabular}} & \multirow{2}{*}{\begin{tabular}[c]{@{}c@{}}SLSIA\\ (SVM)\end{tabular}} & \multirow{2}{*}{\begin{tabular}[c]{@{}c@{}}Avg Loss\end{tabular}} &\multirow{2}{*}{\begin{tabular}[c]{@{}c@{}}Min Loss\\ Time\end{tabular}} &\multirow{2}{*}{\begin{tabular}[c]{@{}c@{}}SLSIA\\ (CNN)\end{tabular}} & \multirow{2}{*}{\begin{tabular}[c]{@{}c@{}}SLSIA\\ (SVM)\end{tabular}}& \multirow{2}{*}{\begin{tabular}[c]{@{}c@{}}Avg Loss\end{tabular}} &\multirow{2}{*}{\begin{tabular}[c]{@{}c@{}}Min Loss\\ Time\end{tabular}} &\multirow{2}{*}{\begin{tabular}[c]{@{}c@{}}SLSIA\\ (CNN)\end{tabular}} & \multirow{2}{*}{\begin{tabular}[c]{@{}c@{}}SLSIA\\ (SVM)\end{tabular}} & \multirow{2}{*}{\begin{tabular}[c]{@{}c@{}}Avg Loss\end{tabular}} &\multirow{2}{*}{\begin{tabular}[c]{@{}c@{}}Min Loss\\ Time\end{tabular}} &\multirow{2}{*}{\begin{tabular}[c]{@{}c@{}}SLSIA\\ (CNN)\end{tabular}} & \multirow{2}{*}{\begin{tabular}[c]{@{}c@{}}SLSIA\\ (SVM)\end{tabular}} \\ 
 & & &&&&&&&&&&&&&&\\
\hline
\addlinespace 
FEMNIST&0.424 &0.372 &0.504 &0.484 &0.424&0.372&0.375&0.356&0.424&0.372&0.304&0.344&0.364&0.332&0.048&0.071\\
Shakespeare&0.452 &0.340 &0.638 &0.648 &0.452&0.340&0.628&0.578&0.452&0.340&0.720&0.584&0.432&0.280&0.622&0.483\\
Synthetic&0.480 &0.360 &0.888&0.868&0.480&0.360&0.911&0.937&0.480&0.360&0.876&0.760&0.460&0.360&0.870&0.800\\

\hline
\end{tabular}
\end{adjustbox}
\end{table*}






\subsection{Analysis}
\label{sub:subject_analysis}

To better understand the attack performance of various methods on different datasets, we analyze embeddings obtained from the pre-trained and local models and the input feature distance between the data from the target subject and other random subjects.

\begin{figure}[!ht]
    \centering
    \begin{subfigure}[b]{0.22\textwidth}
        \centering
        \includegraphics[scale=0.35]{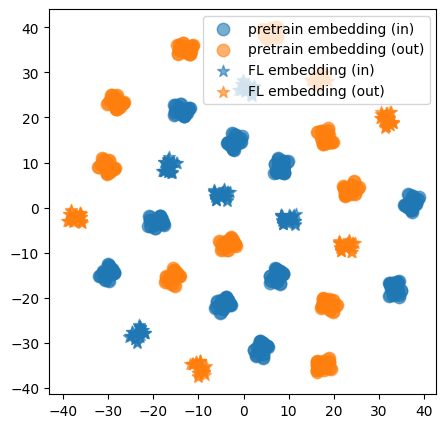}
        \caption{FEMNIST}
        \label{FEMNIST_2-D}
    \end{subfigure}
    \begin{subfigure}[b]{0.22\textwidth}
        \centering
        \includegraphics[scale=0.35]{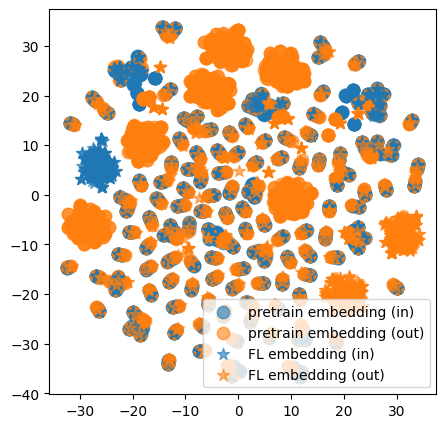}
        \caption{Shakespeare}
        \label{Shakespeare_2-D}
    \end{subfigure}
    \begin{subfigure}[b]{0.22\textwidth}
        \centering
        \includegraphics[scale=0.35]{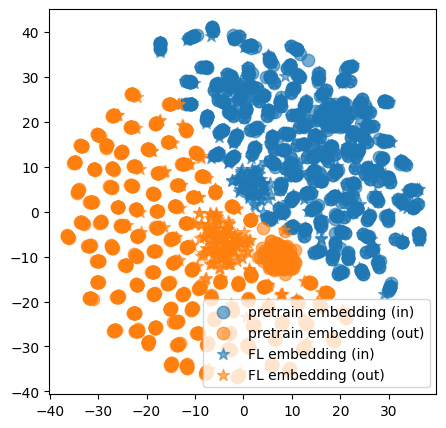}
        \caption{Synthetic}
        \label{Synthetic_2-D}
    \end{subfigure}
    \caption{High-dimensional embeddings projected into 2-dimensional space with t-SNE. We obtain the high-dimensional embeddings by evaluating the evaluation data points from the target subject to pre-trained and local models. Among four sets of embeddings, "pretrain embedding (in)," "pretrain embedding (out)," "FL embedding (in)," and "FL embedding (out)" are from target pre-trained, random pre-trained, target local, and random local models separately.}
    \label{h-D_embedding_to_2-D}
\end{figure}

The "in" and "out" embeddings obtained from the pre-trained and local models expose the transferability of the binary attack model from the training to the evaluation. Figure~\ref{h-D_embedding_to_2-D} shows the projection of the high-dimensional embedding to 2-dimensional points with t-SNE~\cite{wattenberg2016how}. Looking at the results for the Synthetic dataset, Figure~\ref{Synthetic_2-D}, we can see the \textcolor{blue}{separation between the "in" embeddings (yellow) and the "out" embeddings (blue) generated by both pre-trained and local models, which means the binary attack model trained with the embeddings using the embeddings obtained from pre-trained models still performs well on the local models in the Synthetic dataset.}
Looking at the Shakespeare dataset instead, Figure~\ref{Shakespeare_2-D}, there is a partial overlapping area between the "in" and "out" embeddings, and each is characterized by multiple separated clusters rather than a big, more cohesive one, which reduces the transferability of the binary attack model. The "in" and "out" embeddings are divided into more clusters in FEMNIST, Figure~\ref{FEMNIST_2-D}, and even the "in" (or "out") embeddings from pre-trained models are not in the same area. In this case, we can find a classifier to separate the "in" and "out" embeddings from pre-trained models early. However, this model is hard to use in classifying the embeddings from the local clients, which brings low performance to the FEMNIST. 

From Figure~\ref{attack_accuracy_distribution_50subjects}, we observe that the accuracy of the attack varies from close to 0 to 1. To further explore the reason behind accuracy variance, we measure the average input feature distance between the target and other random subjects in the training data of pre-trained and local models. Table~\ref{avg_input_feature_dis} exposes the average input feature distance of the target subject with a high accuracy and the target subject with a low accuracy under each dataset. As expected, the table shows that the target subject with high attack accuracy has input features relatively far from other random subjects, unlike the target subject with relatively low accuracy, which has relatively compact input features among target and random subjects. 
The average input feature distance is the average pair-wise Euclidean distance between two subjects' flattened data input features. For example, the target subject with an accuracy of 0.8 in FEMNIST has average input feature distances of 5.413, 5.472, 5.383, and 5.340, which are about 8\% higher than the target subject with an accuracy of 0.1 with distances of 5.018, 5.065, 4.780, and 4.738. Other datasets have a similar phenomenon with different input feature distance gap levels. We note that Shakespeare's average input feature distance is enormous as we use the index of words in the dictionary as the input feature rather than the embedding of each word after the embedding layer. It will not change the comparison of high and low-accuracy target subjects, as the embedding for the same word is the same. From the above analysis, we conclude that the target subjects, separate from other random subjects, are more vulnerable to our SLSIA. By comparing the average actual (rather than the dictionary index in Shakespeare) input feature distance between the FEMNIST and Synthetic datasets and the attack performance in Table~\ref{avg_metric_value_over_50_subejcts}, we can also observe the dataset with a relatively higher average input feature distance will have a relatively higher attack accuracy under our SLSIA. In other words, the dataset with subjects slightly far from each other is more vulnerable. If the subject is relatively far from other subjects, the models could learn separate embeddings for its data points, which makes the binary attack model easier to learn and transfer from the pre-trained models to the local models. \todo{The significance between the different values of distance is not clear to me. For FEMNIST the difference between 5413 and 5018 is about 8\%, can we consider it significant? maybe speaking in terms of percentage is more straightforward to understand compared to absolute values. let me know what you think. Re: I agree. It is more clear to use the percentage.}

\begin{table*}[ht]
\centering
\caption{The averaged input feature distance between the data from the target and random subjects.}
\label{avg_input_feature_dis}
\begin{adjustbox}{max width=0.64\textwidth}
\begin{tabular}{cccccc}
\hline
\addlinespace 
\multirow{3}{*}{Dataset}  & \multirow{3}{*}{\begin{tabular}[c]{@{}c@{}}SLSIA\\ (CNN) \\ accuracy\end{tabular}} & \multicolumn{4}{c}{the source of random subjects} \\ 
\cmidrule(lr){3-6}
 &  & \multirow{2}{*}{\begin{tabular}[c]{@{}c@{}}random pre-trained\\models\end{tabular}} &\multirow{2}{*}{\begin{tabular}[c]{@{}c@{}}target pre-trained\\models\end{tabular}} &\multirow{2}{*}{\begin{tabular}[c]{@{}c@{}}random local \\ models\end{tabular}} & \multirow{2}{*}{\begin{tabular}[c]{@{}c@{}}target local \\models\end{tabular}}\\ 
 & & &&&\\
\hline
\addlinespace 
\multirow{2}{*}{FEMNIST}&0.8 &5.413 &5.472 &5.383 &5.340\\
&0.1 &5.018 &5.065 &4.780 &4.738\\
\hline
\addlinespace 
\multirow{2}{*}{Shakespeare}&0.9 &71881.506 & 72077.575 &71604.278 &71579.257\\
&0.4 &69823.702 &70085.036 &69905.794 &69608.564\\
\hline
\addlinespace 
\multirow{2}{*}{Synthetic}&1.0 &14.642 &16.464&15.921&18.239\\
&0.6 &14.044 &15.689&15.298&17.202\\
\hline
\end{tabular}
\end{adjustbox}
\end{table*}


\subsection{Ablation Study}
\label{sub:ablation_study}
To explore the factors influencing our SLSIA, we selected ten subjects with relatively high performance from the 50 subjects mentioned before for attacking under different conditions. Due to the space limitation, we show the average accuracy as it can reflect the performance alteration. \textcolor{blue}{This ablation study is structured in three parts: (1) we change the percentage of the target subject's data points in the local client, (2) we test different layers for the generation of the embeddings in input to our attack model, and (3) we test different numbers of local training epochs for both pre-trained and local models.}

(1) \textbf{Changing the percentage of the target subject's data points in the local client}. \textcolor{blue}{In the default setting, we assign a dataset to each target local client, with half of the data points from the target subject and the other half from a random subject.}
Therefore, we change this rate from 50\% to 30\% and 10\% by increasing the number of random subjects and data points from random subjects in each local client and keeping the number of data points from the target subject unchanged. Table~\ref{avg_acc_under_various_rates} shows the average accuracy under various rates. We find that \textbf{Avg Loss} with three datasets and \textbf{Min Loss Time} with FEMNIST and Shakespeare will increase their attack performance while decreasing the rate of the data from the target subject. For example, \textbf{Avg Loss} obtains an accuracy of 0.54, 0.56, and 0.63 under the rate of 50\%, 30\%, and 10\% separately with Synthetic. \textcolor{blue}{As the rate of data points from the target subject decreases, the total number of data points in the training data of the local client increases with the addition of data points from random subjects. More training data points could improve the generalizability of the previous low-performance local model.  We guess that more data leads the local model to learn the target subject better. Hence, the evaluation data points from the target subject ($D_{s_t}^e$) obtain a lower loss on the local clients trained with the data from the target subject, which makes \textbf{Avg Loss} and \textbf{Min Loss Time} a higher attack accuracy.} \todo{This behavior is not clear to me, can you explain better what you meant?} Our SLSIA drops within FEMNIST and Shakespeare while decreasing the rate. However, our SLSIA is still higher than the two previous methods, apart from the Shakespeare under a rate of 10\% (10\% lower in average accuracy). For Synthetic, our SLSIA achieves accuracy of 0.93 and 0.97 even under a rate of 10\%. It indicates that if the subjects are distinguishable enough, our SLSIA could have an average attack accuracy higher than 0.9 with a low rate of data points from the target subject. We attribute the average accuracy drop of our SLSIS with FEMNIST and Shakespeare to the fact that subjects are not distinguishable enough in those two datasets.

\begin{table}[ht]
\centering
\caption{The averaged accuracy of 10 runs (subjects) with various rates of data from the target subject in the target local client.}
\label{avg_acc_under_various_rates}
\begin{adjustbox}{max width=0.48\textwidth}
\begin{tabular}{cccccccccc}
\hline
\addlinespace 
\multirow{3}{*}{Methods}  & \multicolumn{9}{c}{the rate of data from target subject in the target local client}\\ 
\cmidrule(lr){2-10}
& \multicolumn{3}{c}{50\%} & \multicolumn{3}{c}{30\%} & \multicolumn{3}{c}{10\%} \\ 
\cmidrule(lr){2-4} \cmidrule(lr){5-7} \cmidrule(lr){8-10}
 & \rotatebox{90}{FEMNIST} &\rotatebox{90}{Shakespeare} & \rotatebox{90}{Synthetic} & \rotatebox{90}{FEMNIST} &\rotatebox{90}{Shakespeare} & \rotatebox{90}{Synthetic}& \rotatebox{90}{FEMNIST} &\rotatebox{90}{Shakespeare} & \rotatebox{90}{Synthetic} \\ 
\hline
\addlinespace 
Avg Loss&0.30 &0.40&0.54&0.42&0.52&0.56&0.42&0.66& 0.63   \\
\addlinespace
\multirow{2}{*}{\begin{tabular}[c]{@{}c@{}}Min Loss\\ Time\end{tabular}} &\multirow{2}{*}{0.36} &\multirow{2}{*}{0.30}&\multirow{2}{*}{0.40}&\multirow{2}{*}{0.44}&\multirow{2}{*}{0.46}&\multirow{2}{*}{0.36}&\multirow{2}{*}{0.42}&\multirow{2}{*}{0.58}& \multirow{2}{*}{0.30}\\
 & & &&&&&&&\\
\multirow{2}{*}{\begin{tabular}[c]{@{}c@{}}SLSIA\\ (CNN)\end{tabular}} &\multirow{2}{*}{0.67} &\multirow{2}{*}{0.68}&\multirow{2}{*}{0.91}&\multirow{2}{*}{0.62}&\multirow{2}{*}{0.61}&\multirow{2}{*}{0.88}&\multirow{2}{*}{0.48}&\multirow{2}{*}{0.54}&\multirow{2}{*}{0.93} \\
 & & &&&&&&&\\
\multirow{2}{*}{\begin{tabular}[c]{@{}c@{}}SLSIA\\ (SVM)\end{tabular}} &\multirow{2}{*}{0.62} &\multirow{2}{*}{0.70}&\multirow{2}{*}{0.88}&\multirow{2}{*}{0.52}&\multirow{2}{*}{0.62}&\multirow{2}{*}{0.87}&\multirow{2}{*}{0.53}&\multirow{2}{*}{0.53}&\multirow{2}{*}{0.97} \\
 & & &&&&&&&\\
\hline
\end{tabular}
\end{adjustbox}
\end{table}


(2)~\label{different_layer_exploration} \textbf{Different layers of the target (pre-trained or local) model as embeddings}. 
\textcolor{blue}{The embeddings in input to our attack model can be generated from different layers of target models. In this study, we explore the effectiveness of our attack while generating the embeddings from different layers.}
By default, we extract the embeddings from layer 1 in CNN (the flattened output of the second convolutional layer), layer 0 in LSTM (the last hidden state of the LSTM layer), and layer 0 in MLP (the output of the first linear layer). From Table~\ref{accuracy_over_different_layers}, we can see how the attack accuracy of three target subjects from three datasets with different layers changes. As we can see, our selected layers have relatively higher performance (0.8, 0.8, and 1.0 with CNN-based SLSIA) than the others. This indicates that the early layers of the model, rather than the last two layers (logits and confidence scores) of each model in the table, will expose more information about the data from the target subject. This corresponds to the finding that the early layer will learn easy but distinguishable features while the higher layer will obtain abstract features~\cite{mahendran2015understanding,Christian_properties_2014}.\todo{citation} The early layer is beneficial to our SLSIA.

\begin{table*}[ht]
\centering
\caption{The attack accuracy of one run (subject) with different layers under each type of model.}
\label{accuracy_over_different_layers}
\begin{adjustbox}{max width=0.88\textwidth}
\begin{tabular}{ccccccccccccc}
\hline
\addlinespace 
\multirow{2}{*}{\begin{tabular}[c]{@{}c@{}} SLSIA \\ methods\end{tabular}}  & \multicolumn{6}{c}{CNN (FEMNIST) layer index} & \multicolumn{3}{c}{LSTM (Shakespeare) layer index} &\multicolumn{3}{c}{MLP (Synthetic) layer index} \\ 
\cmidrule(lr){2-7} \cmidrule(lr){8-10} \cmidrule(lr){11-13}
& layer 0 & layer 1& layer 2 & layer 3 & layer 4 & layer 5 & layer 0 & layer 1 & layer 2 & layer 0 & layer 1 & layer 2\\
\hline
\addlinespace 
\multirow{2}{*}{\begin{tabular}[c]{@{}c@{}}SLSIA\\ (CNN)\end{tabular}} &\multirow{2}{*}{0.8} &\multirow{2}{*}{0.8} &\multirow{2}{*}{0.6} &\multirow{2}{*}{0.5} &\multirow{2}{*}{0.6} &\multirow{2}{*}{0.5} &\multirow{2}{*}{0.8}&\multirow{2}{*}{0.8}&\multirow{2}{*}{0.5} &\multirow{2}{*}{1.0}&\multirow{2}{*}{0.1}&\multirow{2}{*}{0.2}\\
 & & & & & & &&& &&&\\
\multirow{2}{*}{\begin{tabular}[c]{@{}c@{}}SLSIA\\ (SVM)\end{tabular}}&\multirow{2}{*}{0.5} &\multirow{2}{*}{0.8} &\multirow{2}{*}{0.7} &\multirow{2}{*}{0.8} &\multirow{2}{*}{0.8} & \multirow{2}{*}{0.8} &\multirow{2}{*}{0.8}&\multirow{2}{*}{0.8}&\multirow{2}{*}{0.6} &\multirow{2}{*}{1.0}&\multirow{2}{*}{0.1}&\multirow{2}{*}{0.1}\\
 & & & & & & &&& &&&\\
\hline
\end{tabular}
\end{adjustbox}
\end{table*}


(3) \textbf{The local training epochs in pertaining and actual FL}. As we mentioned, the low performance of the \textbf{Avg Loss} and \textbf{Min Loss Time} methods might be related to the inefficient learning of the model at the first FL round in Section~\ref{sub:attackPerfromance}. Therefore, we increase the local epoch of the first round to observe the performance alteration in Table~\ref{avg_acc_with_various_local_epochs}. \textcolor{blue}{We find a similar rule as the exploration of decreasing the rate of data points from the target subject in local clients.} \textbf{Avg Loss} and \textbf{Min Loss Time} mainly increase their performance with more local training epochs apart from the Synthetic under \textbf{Min Loss Time}. For example, \textbf{Avg Loss} obtains an accuracy of 0.40, 0.54, and 0.78 under the local training epochs of 5, 10, and 15 separately within Synthetic. It indicates more local training epochs at the first FL round increase the performance of the two previous attacking methods, which proves our guess that better learning of the local models will increase their attack performance. \textcolor{blue}{Instead of improving the generalizability of the local model via more data points, increasing the local training epoch is also a way to make the local model learn better about the target subject, which reduces the loss of the data points from the target subject and increases the attack performance of \textbf{Avg Loss} and \textbf{Min Loss Time} as they rely on the low loss of the local client on data points from the target subject.} Even though our SLSIA drops with more local training epochs under the FEMNIST and Shakespeare, our SLSIA is still higher than the previous two methods. In Synthetic, our SLSIA obtains an accuracy of over 0.9 under different local training epochs, which also indicates the effectiveness of our SLSIA. \todo{As the first study, this behavior is not clear to me, can you explain better what you meant?}

\begin{table}[ht]
\centering
\caption{The averaged accuracy of 10 runs (subjects) with various local training epochs.}
\label{avg_acc_with_various_local_epochs}
\begin{adjustbox}{max width=0.48\textwidth}
\begin{tabular}{cccccccccc}
\hline
\addlinespace 
\multirow{3}{*}{Methods}  & \multicolumn{9}{c}{local training epochs (pretraining epochs)}\\ 
\cmidrule(lr){2-10}
& \multicolumn{3}{c}{5} & \multicolumn{3}{c}{10} & \multicolumn{3}{c}{15} \\ 
\cmidrule(lr){2-4} \cmidrule(lr){5-7} \cmidrule(lr){8-10}
 & \rotatebox{90}{FEMNIST} &\rotatebox{90}{Shakespeare} & \rotatebox{90}{Synthetic} & \rotatebox{90}{FEMNIST} &\rotatebox{90}{Shakespeare} & \rotatebox{90}{Synthetic}& \rotatebox{90}{FEMNIST} &\rotatebox{90}{Shakespeare} & \rotatebox{90}{Synthetic} \\ 
\hline
\addlinespace 
Avg Loss&0.30 &0.40&0.54&0.26&0.54&0.54&0.40&0.78& 0.61  \\
\addlinespace
\multirow{2}{*}{\begin{tabular}[c]{@{}c@{}}Min Loss\\ Time\end{tabular}} &\multirow{2}{*}{0.36} &\multirow{2}{*}{0.30}&\multirow{2}{*}{0.40}&\multirow{2}{*}{0.36}&\multirow{2}{*}{0.42}&\multirow{2}{*}{0.30}&\multirow{2}{*}{0.50}&\multirow{2}{*}{0.42}& \multirow{2}{*}{0.24}\\
 & & &&&&&&&\\
\multirow{2}{*}{\begin{tabular}[c]{@{}c@{}}SLSIA\\ (CNN)\end{tabular}} &\multirow{2}{*}{0.67} &\multirow{2}{*}{0.68}&\multirow{2}{*}{0.91}&\multirow{2}{*}{0.52}&\multirow{2}{*}{0.61}&\multirow{2}{*}{0.96}&\multirow{2}{*}{0.53}&\multirow{2}{*}{0.60}&\multirow{2}{*}{0.95} \\
 & & &&&&&&&\\
\multirow{2}{*}{\begin{tabular}[c]{@{}c@{}}SLSIA\\ (SVM)\end{tabular}} &\multirow{2}{*}{0.62} &\multirow{2}{*}{0.70}&\multirow{2}{*}{0.88}&\multirow{2}{*}{0.57}&\multirow{2}{*}{0.59}&\multirow{2}{*}{0.98}&\multirow{2}{*}{0.52}&\multirow{2}{*}{0.67}&\multirow{2}{*}{0.95} \\
 & & &&&&&&&\\
\hline
\end{tabular}
\end{adjustbox}
\end{table}

\subsection{Defense}
\label{sub:defense}

From the previous analysis, we conclude that the subject that is relatively far from other subjects is vulnerable under our SLSIA. \textcolor{blue}{This finding indicates that training the local models with vulnerable subjects makes them produce more distinguishable embeddings on data from the same vulnerable source.}
To eliminate the contribution of data from vulnerable subjects, we employ subject-level differential privacy~\cite{marathe2022subject}, which regards data from a subject as a data point in item-level differential privacy~\cite{Abadi_2016_DPML}, to protect the data from each subject. In the previous work~\cite{suri2022subject}, they apply the Hierarchical Gradient Averaging~\cite{marathe2022subject} to achieve subject-level differential privacy. We follow their work to implement subject-level differential privacy using the same method and compare it with item-level differential privacy.

We only apply item-level and subject-level differential privacy in the training of each local client.
\textcolor{blue}{Since, in our case, the attacker is an honest but curious server, instead of the user-level differential privacy as in FL averaging~\cite{brendan2018learning}, we have to defend the local models of the clients rather than the global one to follow the scenario presented by Hu et al.~\cite{SIA_Hu}. }
Unlike the non-private training, applying differential privacy requires clipping the per-sample gradient and adding noise to the averaged gradient. Hence, we utilize the same standard hyper-parameters in non-private training for private training, including the expected batch size of 12, local epochs of 5, and a learning rate of 0.01. \textcolor{blue}{The clipping threshold ($C$) and the noise multiplier ($\sigma$) are two parameters utilized for applying differential privacy in deep learning~\cite{Abadi_2016_DPML}. The clipping threshold determines the maximum gradient of each training data point. If the gradient of the data point is larger than $C$, the differential privacy algorithm will reduce it to $C$. The noise multiplier is related to the Gaussian noise added to the gradient.}\todo{this sentence is not clear, can you please rephrase it} Once those two parameters are determined, we can formulate the privacy cost ($\epsilon$ and $\delta$) of the local client along the local training epoch. It means we obtain the privacy budget $\epsilon$ under a certain value of $\delta$. Following the previous works~\cite{suri2022subject, Abadi_2016_DPML}, we set $\delta=10^{-5}$ to calculate privacy budget $\epsilon$. 
\textcolor{blue}{In our experiments, for the clipping parameter, we set C=1.0 as the default configuration in Opacus~\cite{opacus_2024} to follow the previous work~\cite{carlini_membership_2021}. We set $\sigma=0.5$ for the noise multiplier as in the previous work~\cite{loss_trajectory_liu}.}
Due to the variance of the number of data points in the training data of local clients and an identical expected batch size, the privacy budget $\epsilon$ of local models varies, which leads us to average the privacy budget of local models obtained from 10 runs (subjects). Besides, we expose the average accuracy change of the original task with the evaluation ('test' column name in the table) and training ('train' column name in the table) data points from the target subject on the target local clients. Finally, we exhibit the performance alteration of our \textbf{SLSIA (CNN)} with differential privacy.

Table~\ref{subject-level_dp} results from applying subject-level differential privacy, and Table~\ref{item-level_dp} results from applying item-level differential privacy. Subject-level differential privacy closes the training and test accuracy to a random guess. For the Synthetic dataset, the drop of 20\% in the training accuracy causes a 36\% decrease in the average accuracy of \textbf{SLSIA (CNN)}. As expected, the reduction of attack performance is related to a significant accuracy drop in the original task. For the Shakespeare dataset, the attack accuracy is only a 4\% drop even though the accuracy of the original task decreases by more than 1000X, indicating the current extent of subject-level differential privacy can not prevent our SLSIA completely. While applying item-level differential privacy with a similar level of privacy budget, the accuracy change of the original task is slight apart from a 20\% drop in the training accuracy of the Synthetic dataset, and the attack accuracy of \textbf{SLSIA (CNN)} has no apparent drop with Shakespeare and even slightly increase under FEMNIST. For Synthetic, the attack accuracy has a drop of more than 30\% to a random guess. The Synthetic dataset is sensitive to item-level and subject-level differential privacy, while the other two are slightly defended under item-level differential privacy. It indicates that subject-level differential privacy is more potent than item-level differential privacy to defend both SMIA~\cite{suri2022subject} and our SLSIA, \textcolor{blue}{while still achieving a higher attack accuracy concerning the compared attacks, without defense, presented in Section~\ref{sub:attackPerfromance}.}

\begin{table*}[ht]
\centering
\caption{The average privacy budget, average accuracy change, and average attack performance alteration of 10 runs with subject-level differential privacy.}
\label{subject-level_dp}
\begin{adjustbox}{max width=0.68\textwidth}
\begin{tabular}{cccccccc}
\hline
\addlinespace 
\multirow{3}{*}{Dataset}  & \multicolumn{3}{c}{non-private} & \multicolumn{4}{c}{$\sigma=0.5$, $C=1.0$, $\delta=10^{-5}$} \\ 
\cmidrule(lr){2-4} \cmidrule(lr){5-8}
 &\multicolumn{2}{c}{Avg task accuracy} &  \multirow{2}{*}{\begin{tabular}[c]{@{}c@{}}Avg SLSIA (CNN)\\accuracy\end{tabular}}  & \multirow{2}{*}{Avg $\epsilon$} &\multicolumn{2}{c}{Avg task accuracy} &  \multirow{2}{*}{\begin{tabular}[c]{@{}c@{}}Avg SLSIA (CNN)\\accuracy\end{tabular}} \\ 
\cmidrule(lr){2-3} \cmidrule(lr){6-7}
 & train & test & & & train & test &\\
\hline
\addlinespace 
FEMNIST&17.6\% &8\% & 0.61 & 30.6 & 11.6\% & 11.5\%& 0.48\\

Shakespeare&4.7\% & 2.8\% & 0.73 & 15.0 &0.002\% &0.009\%& 0.68\\
 
Synthetic&74.8\% &50.0\% &0.89&22.0&54.9\%&50.4\% &0.53\\
\hline
\end{tabular}
\end{adjustbox}
\end{table*}

\begin{table*}[ht]
\centering
\caption{The average privacy budget, average accuracy change, and average attack performance alteration of 10 runs with item-level differential privacy.}
\label{item-level_dp}
\begin{adjustbox}{max width=0.68\textwidth}
\begin{tabular}{cccccccc}
\hline
\addlinespace 
\multirow{3}{*}{Dataset}  & \multicolumn{3}{c}{non-private} & \multicolumn{4}{c}{$\sigma=0.5$, $C=1.0$, $\delta=10^{-5}$} \\ 
\cmidrule(lr){2-4} \cmidrule(lr){5-8}
 &\multicolumn{2}{c}{Avg task accuracy} &  \multirow{2}{*}{\begin{tabular}[c]{@{}c@{}}Avg SLSIA (CNN)\\accuracy\end{tabular}}  & \multirow{2}{*}{Avg $\epsilon$} &\multicolumn{2}{c}{Avg task accuracy} &  \multirow{2}{*}{\begin{tabular}[c]{@{}c@{}}Avg SLSIA (CNN)\\accuracy\end{tabular}} \\ 
\cmidrule(lr){2-3} \cmidrule(lr){6-7}
 & train & test & & & train & test &\\
\hline
\addlinespace 
FEMNIST&17.7\% &7.1\% &0.56 &30.7 &17.0\%&7.2\% &0.58\\

Shakespeare&5.0\% &3.3\% & 0.62 &15.4 &2.0\%&1.0\% &0.58\\
 
Synthetic&76.5\% &51.4\% &0.88 &22.0&56.5\%&50.9\% &0.54\\
\hline
\end{tabular}
\end{adjustbox}
\end{table*}




\section{Related Works}
\label{sec:RW}

This part reviews recent works related to gradient leakage (Section~\ref{GLA}), reconstruction (Section~\ref{RA}), property inference (Section~\ref{PI}), membership inference (Section~\ref{MI}), and source inference (Section~\ref{SI}) attacks in FL.

\subsection{Gradient Leakage Attack}
\label{GLA}

The gradient leakage attack utilizes the exposed gradient updates from local clients to recover the training data or labels from local clients. Phong et al.~\cite{Phong_2017_HE_on_FL} show an honest but curious server can extract the training data from partial gradients uploaded by the local client as the division between the gradient on the weight and bias is related to one specific input value (like one image pixel). Zhu et al.~\cite{Zhu_2019_Deep_leakage} recover the training data and labels of local clients by the iterative update of dummy inputs and labels to match the gradient update via L-BFGS~\cite{Dong_1989_LBFGS}.
Similarly, in a Vertical Federated Learning configuration, Arazzi et al.~\cite{arazzi2023blindsage} propose a gradients leaking approach exploiting the matching between the partial gradients returned by the server to obtain the private labels unavailable to the passive parties.
Considering the inconsistency of convergence and discovering ground-truth labels in~\cite{Zhu_2019_Deep_leakage}, Zhao et al.~\cite{zhao_2020_idlg} propose a reliable method to extract ground-truth labels from shared gradients via observing the signs of gradients on the parameters of the last layer. Geiping et al.~\cite{Geiping_2020_invert_gradient} utilize Adam~\cite{Adam_Diederik_2015} to optimize a cosine similarity with a regularization term about the total variation of the input to align the gradient updates of original and reconstructed inputs instead of a Euclidean loss optimized with L-BFGS~\cite{Dong_1989_LBFGS} in previous works~\cite{Zhu_2019_Deep_leakage,zhao_2020_idlg}.

\subsection{Reconstruction Attack}
\label{RA}
The reconstruction attack targets reconstructing local clients' training data with information not limited to the uploaded gradients. Usually, it uses the GAN to help the reconstruction process. Hitaj et al.~\cite{Hitaj_GAN_2017} implement a GAN-based reconstruction attack, which generates the images belonging to the label that the adversary (a malicious federation user) does not hold. Specifically, the adversary trains a local GAN to generate images whose label is in a victim user by utilizing the global model as the discriminator and mislabeling the generated images as the label the adversary holds, which makes the victim user reveal more information about the target label of the adversary. Wang et al.~\cite{Wang_2019_User_level} propose an exact data reconstruction against the victim client from the perspective of the central server without disrupting FL via training a GAN with a multi-task discriminator, which distinguishes not only fake and real images but also different categories and identity information (from victim client or other clients). With the GAN, the server can generate data (labeled with a specific category) from the victim client.

\subsection{Property Inference Attack}
\label{PI}
The property inference attack aims to infer the private properties of the whole training data of the model. 
Ateniese et al.~\cite{Ateniese_2015_Hacking} extract features (e.g., the support vectors of SVM) from the target model to predict the property of its training data via training a set of shadow classifiers with and without the property. For neural networks, Ganju et al.~\cite{Ganju_2018_PIA} take a neural network's parameters as the feature representation. Zhang et al.~\cite{zhang2021leakage} query the target model and concatenate the posterior probability vectors of data points from an attack dataset to construct the feature for the property inference. Mahloujifar et al.~\cite{Mahloujifar_2022_PIA_from_poisoning} explore whether the poisoning attack will enlarge the information leakage measured with the property inference attack. Wang et al.~\cite{Wang_2023_PAPIA} infer the property of the global model with the help of poisoned data used in adversarial local clients. Suri et al.~\cite{suri2022formalizing} formulate the property and attribute attacks as the distribution inference attack and extend the meta-classifier in the work of Ganju et al.~\cite{Ganju_2018_PIA} to CNNs. In their subsequent work~\cite{Suri_2023_DIA}, they propose a black-box property inference attack by comparing the probability distribution with the KL-divergence value. Hartmann et al.~\cite{Hartmann_2023_DIA} analyze the reasons for distribution leakage. 

\subsection{Membership Inference Attack}
\label{MI}
The membership inference attack infers the existence of a single data point or a group of data points belonging to a subject in the model's training data. Melis et al.~\cite{melis_2019_exploiting} infer the membership of a text or location in other clients by checking whether the embedding weights of its tokens are all non-zero in the aggregated embedding layer. Nasr et al.~\cite{nasr_2019_comprehensive} utilize the gradients of all parameters on the target data point as the attack feature. Zari et al.~\cite{zari_2021_efficient} reduce the heavy feature of the attack model in~\cite{nasr_2019_comprehensive} by combining the outputs of models among training rounds. Gu et al.~\cite{Gu_2022_CSMIA} leverage the prediction confidence, measured with the prediction probability of the ground truth label and the modified prediction entropy~\cite{song_systematic_2021} for membership inference in FL. Li et al.~\cite{li2023effective} calculate the cosine similarity between the gradient of the target data point and the sum of gradients to predict the membership. Zhang et al.~\cite{Zhang_2022_GANMIA} train a GAN to generate data points close to the training data of the target FL model for membership inference. Chen et al.~\cite{Chen_user-level_2020} infer whether the data is from one victim client from the perspective of another client, assuming all clients do not share identical labels. In their extensive work~\cite{Zhao_user-level_2021}, they isolate the updated model from the victim client and train a GAN to generate data points for the membership inference. Zhang et al.~\cite{Zhang_2023_efficient_MIA} propose to use the model's biases rather than weights for the membership inference attack. Pichler et al.~\cite{pichler2022perfectly} implement a perfectly accurate membership inference attack by inserting an intentionally designed network into the local client and observing a bias modification magnitude.

Suri et al.~\cite{suri2022subject} propose the SMIA to infer whether any client trains its local model with data points from the target subject in cross-silo FL. They put forward the Loss-Threshold and Loss-Across-Rounds attacks. However, they can not precisely distinguish which client trains its local model with the data from the target subject. Liu et al.~\cite{Liu_23_SubjectMIA} investigate the SMIA on a group of concerned clients. However, there are some strong assumptions about their work. The first is that the adversary holds 10\% of the data points of all subjects in the training data of concerned clients. Second, they can not precisely detect all clients trained with the data from the target subject. Third, the adversary interrupts the FL training by uploading the trained discriminator model to the server. Besides, they do not use the dataset with a clear concept of subject for evaluation but model each subject as a parametric distribution using a multivariate Gaussian distribution.

\subsection{Source Inference Attack}
\label{SI}
The SIA predicts the exact client that trains its model with a single data point or a group of data points from a subject. Hu et al.~\cite{SIA_Hu} propose the SIA with a strategy that calculates the losses of this data point on all clients and predicts the client with the lowest loss on this data point as the client trains with it from the perspective of a curious but honest central serve in FL. However, they assume that only one client trains its local model with the target data point, ignoring the scenarios that none or more than one client prepares local models with this data point, which is infeasible with their strategy. Besides, instead of holding a target data point in clients' training data to investigate the source, the central sever cares more about whether the data from a specific subject is used for local training and where the data is used without knowing which data is used. 


\section{Conclusion and Future Work}
\label{sec:Conclusion}
This paper proposes a novel approach for a source inference attack in Cross-Silo Federated Learning.
Due to its characteristics, our attack can be successfully adopted as a data auditing tool to assess whether data from a specific subject have been used during the FL task.
To develop our solution, we leveraged the early embedding of local models to expose whether their training data contained data points from a specific target subject. Specifically, we audit the subject data usage of local clients by training an attack model to predict whether an embedding associated with a data point is produced by a model originally trained with data from a target subject. The relatively higher attack performance of our SLSIA on three datasets than those of previous methods empirically proves its adequateness for subject data auditing. Analyzing the embeddings and the input feature distance, we found that datasets with more distinguishable subjects are more vulnerable to our SLSIA. The subject-level differential privacy can significantly decrease the accuracy of our SLSIA with a considerable utility drop of the global FL model and, in most cases, cannot prevent our SLSIA completely.

As we assume that the central server will select all local models for updating the global model in the first round, we leave the extension to scenarios where not all local models are utilized at once as one of future work. Besides, a potential direction to improve our solution is identifying strategies to reduce the data from the target subject required for the inference attack. This is especially relevant in scenarios in which the distribution of data from different subjects is similar. In addition, an efficient and less impacting defense mechanism is also needed, as differential privacy greatly compromises the original task performance.

\printcredits

\section*{Declaration of competing interest}
The authors declare no conflict of interest.

\section*{Data availability}
We use three datasets in our experiments. FEMNIST and Shakespeare are from LEAF (\url{https://leaf.cmu.edu/}), an open-source benchmark for Federated Learning. We generate Synthetic according to the experiments we need. This generated dataset is available in a public repository (\url{https://drive.google.com/file/d/1V5rhqQCmSi3kW64F31_O8n-QgAmdRIUR/view?usp=drive_link}).

\section*{Acknowledgement}

This work has been funded by the Italian Ministry of University and Research through the PRIN Project ``HOMEY: a Human-centric IOE-based framework for supporting the transition towards industry 5.0'' (code 2022NX7WKE). Besides, this research was partially supported by the Chinese Scholarship Council (CSC).


\bibliographystyle{cas-model2-names}

\bibliography{bibliography}

\begin{thebibliography}{53}
\expandafter\ifx\csname natexlab\endcsname\relax\def\natexlab#1{#1}\fi
\providecommand{\url}[1]{\texttt{#1}}
\providecommand{\href}[2]{#2}
\providecommand{\path}[1]{#1}
\providecommand{\DOIprefix}{doi:}
\providecommand{\ArXivprefix}{arXiv:}
\providecommand{\URLprefix}{URL: }
\providecommand{\Pubmedprefix}{pmid:}
\providecommand{\doi}[1]{\href{http://dx.doi.org/#1}{\path{#1}}}
\providecommand{\Pubmed}[1]{\href{pmid:#1}{\path{#1}}}
\providecommand{\bibinfo}[2]{#2}
\ifx\xfnm\relax \def\xfnm[#1]{\unskip,\space#1}\fi
\bibitem[{Abadi et~al.(2016)Abadi, Chu, Goodfellow, McMahan, Mironov, Talwar and Zhang}]{Abadi_2016_DPML}
\bibinfo{author}{Abadi, M.}, \bibinfo{author}{Chu, A.}, \bibinfo{author}{Goodfellow, I.}, \bibinfo{author}{McMahan, H.B.}, \bibinfo{author}{Mironov, I.}, \bibinfo{author}{Talwar, K.}, \bibinfo{author}{Zhang, L.}, \bibinfo{year}{2016}.
\newblock \bibinfo{title}{Deep learning with differential privacy}, in: \bibinfo{booktitle}{Proceedings of the 2016 ACM SIGSAC Conference on Computer and Communications Security}, \bibinfo{publisher}{Association for Computing Machinery}. p. \bibinfo{pages}{308–318}.
\newblock \URLprefix \url{https://doi.org/10.1145/2976749.2978318}, \DOIprefix\doi{10.1145/2976749.2978318}.
\bibitem[{Arazzi et~al.(2023)Arazzi, Conti, Koffas, Krcek, Nocera, Picek and Xu}]{arazzi2023blindsage}
\bibinfo{author}{Arazzi, M.}, \bibinfo{author}{Conti, M.}, \bibinfo{author}{Koffas, S.}, \bibinfo{author}{Krcek, M.}, \bibinfo{author}{Nocera, A.}, \bibinfo{author}{Picek, S.}, \bibinfo{author}{Xu, J.}, \bibinfo{year}{2023}.
\newblock \bibinfo{title}{Blindsage: Label inference attacks against node-level vertical federated graph neural networks}.
\newblock \bibinfo{journal}{arXiv preprint arXiv:2308.02465} .
\bibitem[{Ateniese et~al.(2015)Ateniese, Mancini, Spognardi, Villani, Vitali and Felici}]{Ateniese_2015_Hacking}
\bibinfo{author}{Ateniese, G.}, \bibinfo{author}{Mancini, L.V.}, \bibinfo{author}{Spognardi, A.}, \bibinfo{author}{Villani, A.}, \bibinfo{author}{Vitali, D.}, \bibinfo{author}{Felici, G.}, \bibinfo{year}{2015}.
\newblock \bibinfo{title}{Hacking smart machines with smarter ones: How to extract meaningful data from machine learning classifiers}.
\newblock \bibinfo{journal}{Int. J. Secur. Netw.} \bibinfo{volume}{10}, \bibinfo{pages}{137–150}.
\newblock \URLprefix \url{https://doi.org/10.1504/IJSN.2015.071829}, \DOIprefix\doi{10.1504/IJSN.2015.071829}.
\bibitem[{BUKATY(2019)}]{CCPA_citation}
\bibinfo{author}{BUKATY, P.}, \bibinfo{year}{2019}.
\newblock \bibinfo{title}{The California Consumer Privacy Act (CCPA): An implementation guide}.
\newblock \bibinfo{publisher}{IT Governance Publishing}.
\newblock \URLprefix \url{http://www.jstor.org/stable/j.ctvjghvnn}.
\bibitem[{Caldas et~al.(2018)Caldas, Duddu, Wu, Li, Kone{\v{c}}n{\`y}, McMahan, Smith and Talwalkar}]{caldas2018leaf}
\bibinfo{author}{Caldas, S.}, \bibinfo{author}{Duddu, S.M.K.}, \bibinfo{author}{Wu, P.}, \bibinfo{author}{Li, T.}, \bibinfo{author}{Kone{\v{c}}n{\`y}, J.}, \bibinfo{author}{McMahan, H.B.}, \bibinfo{author}{Smith, V.}, \bibinfo{author}{Talwalkar, A.}, \bibinfo{year}{2018}.
\newblock \bibinfo{title}{Leaf: A benchmark for federated settings}.
\newblock \bibinfo{journal}{arXiv preprint arXiv:1812.01097} .
\bibitem[{Carlini et~al.(2022)Carlini, Chien, Nasr, Song, Terzis and Tram{\`{e}}r}]{carlini_membership_2021}
\bibinfo{author}{Carlini, N.}, \bibinfo{author}{Chien, S.}, \bibinfo{author}{Nasr, M.}, \bibinfo{author}{Song, S.}, \bibinfo{author}{Terzis, A.}, \bibinfo{author}{Tram{\`{e}}r, F.}, \bibinfo{year}{2022}.
\newblock \bibinfo{title}{Membership inference attacks from first principles}, in: \bibinfo{booktitle}{Proceedings of the 43rd {IEEE} Symposium on Security and Privacy}, \bibinfo{publisher}{IEEE Computer Society}. pp. \bibinfo{pages}{1897--1914}.
\newblock \URLprefix \url{https://doi.org/10.1109/SP46214.2022.9833649}, \DOIprefix\doi{10.1109/SP46214.2022.9833649}.
\bibitem[{Chen et~al.(2020)Chen, Zhang, Zhao, Han, Zhu and Chen}]{Chen_user-level_2020}
\bibinfo{author}{Chen, J.}, \bibinfo{author}{Zhang, J.}, \bibinfo{author}{Zhao, Y.}, \bibinfo{author}{Han, H.}, \bibinfo{author}{Zhu, K.}, \bibinfo{author}{Chen, B.}, \bibinfo{year}{2020}.
\newblock \bibinfo{title}{Beyond model-level membership privacy leakage: an adversarial approach in federated learning}, in: \bibinfo{booktitle}{2020 29th International Conference on Computer Communications and Networks (ICCCN)}, \bibinfo{publisher}{{IEEE} Computer Society}. pp. \bibinfo{pages}{1--9}.
\newblock \DOIprefix\doi{10.1109/ICCCN49398.2020.9209744}.
\bibitem[{Contributors()}]{pytorch_lstm_2024}
\bibinfo{author}{Contributors, P.}, .
\newblock \bibinfo{title}{Pytorch lstm}.
\newblock \bibinfo{howpublished}{\url{https://pytorch.org/docs/stable/generated/torch.nn.LSTM.html}}.
\newblock \bibinfo{note}{Accessed: 2024-05-27}.
\bibitem[{Deng(2012)}]{Deng_2012_MNIST}
\bibinfo{author}{Deng, L.}, \bibinfo{year}{2012}.
\newblock \bibinfo{title}{The mnist database of handwritten digit images for machine learning research [best of the web]}.
\newblock \bibinfo{journal}{IEEE Signal Processing Magazine} \bibinfo{volume}{29}, \bibinfo{pages}{141--142}.
\newblock \DOIprefix\doi{10.1109/MSP.2012.2211477}.
\bibitem[{scikit-learn developers()}]{sklearn_svc_2024}
\bibinfo{author}{scikit-learn developers}, .
\newblock \bibinfo{title}{Sklearn svm}.
\newblock \bibinfo{howpublished}{\url{https://scikit-learn.org/stable/modules/generated/sklearn.svm.SVC.html}}.
\newblock \bibinfo{note}{Accessed: 2024-05-27}.
\bibitem[{Ganju et~al.(2018)Ganju, Wang, Yang, Gunter and Borisov}]{Ganju_2018_PIA}
\bibinfo{author}{Ganju, K.}, \bibinfo{author}{Wang, Q.}, \bibinfo{author}{Yang, W.}, \bibinfo{author}{Gunter, C.A.}, \bibinfo{author}{Borisov, N.}, \bibinfo{year}{2018}.
\newblock \bibinfo{title}{Property inference attacks on fully connected neural networks using permutation invariant representations}, in: \bibinfo{booktitle}{Proceedings of the 2018 ACM SIGSAC Conference on Computer and Communications Security}, \bibinfo{publisher}{Association for Computing Machinery}. p. \bibinfo{pages}{619–633}.
\newblock \URLprefix \url{https://doi.org/10.1145/3243734.3243834}, \DOIprefix\doi{10.1145/3243734.3243834}.
\bibitem[{Geiping et~al.(2020)Geiping, Bauermeister, Dr\"{o}ge and Moeller}]{Geiping_2020_invert_gradient}
\bibinfo{author}{Geiping, J.}, \bibinfo{author}{Bauermeister, H.}, \bibinfo{author}{Dr\"{o}ge, H.}, \bibinfo{author}{Moeller, M.}, \bibinfo{year}{2020}.
\newblock \bibinfo{title}{Inverting gradients - how easy is it to break privacy in federated learning?}, in: \bibinfo{booktitle}{Proceedings of the 34th International Conference on Neural Information Processing Systems}, \bibinfo{publisher}{Curran Associates Inc.}
\bibitem[{Gu et~al.(2022)Gu, Bai and Xu}]{Gu_2022_CSMIA}
\bibinfo{author}{Gu, Y.}, \bibinfo{author}{Bai, Y.}, \bibinfo{author}{Xu, S.}, \bibinfo{year}{2022}.
\newblock \bibinfo{title}{Cs-mia: Membership inference attack based on prediction confidence series in federated learning}.
\newblock \bibinfo{journal}{Journal of Information Security and Applications} \bibinfo{volume}{67}, \bibinfo{pages}{103201}.
\newblock \URLprefix \url{https://www.sciencedirect.com/science/article/pii/S2214212622000801}, \DOIprefix\doi{https://doi.org/10.1016/j.jisa.2022.103201}.
\bibitem[{Hartmann et~al.(2023)Hartmann, Meynent, Peyrard, Dimitriadis, Tople and West}]{Hartmann_2023_DIA}
\bibinfo{author}{Hartmann, V.}, \bibinfo{author}{Meynent, L.}, \bibinfo{author}{Peyrard, M.}, \bibinfo{author}{Dimitriadis, D.}, \bibinfo{author}{Tople, S.}, \bibinfo{author}{West, R.}, \bibinfo{year}{2023}.
\newblock \bibinfo{title}{Distribution inference risks: Identifying and mitigating sources of leakage}, in: \bibinfo{booktitle}{2023 IEEE Conference on Secure and Trustworthy Machine Learning (SaTML)}, \bibinfo{publisher}{IEEE Computer Society}. pp. \bibinfo{pages}{136--149}.
\newblock \URLprefix \url{https://doi.ieeecomputersociety.org/10.1109/SaTML54575.2023.00018}, \DOIprefix\doi{10.1109/SaTML54575.2023.00018}.
\bibitem[{Hitaj et~al.(2017)Hitaj, Ateniese and Perez-Cruz}]{Hitaj_GAN_2017}
\bibinfo{author}{Hitaj, B.}, \bibinfo{author}{Ateniese, G.}, \bibinfo{author}{Perez-Cruz, F.}, \bibinfo{year}{2017}.
\newblock \bibinfo{title}{Deep models under the gan: Information leakage from collaborative deep learning}, in: \bibinfo{booktitle}{Proceedings of the 2017 ACM SIGSAC Conference on Computer and Communications Security}, \bibinfo{publisher}{Association for Computing Machinery}. p. \bibinfo{pages}{603–618}.
\newblock \URLprefix \url{https://doi.org/10.1145/3133956.3134012}, \DOIprefix\doi{10.1145/3133956.3134012}.
\bibitem[{Hu et~al.(2023)Hu, Zhang, Salcic, Sun, Choo and Dobbie}]{SIA_Hu}
\bibinfo{author}{Hu, H.}, \bibinfo{author}{Zhang, X.}, \bibinfo{author}{Salcic, Z.}, \bibinfo{author}{Sun, L.}, \bibinfo{author}{Choo, K.}, \bibinfo{author}{Dobbie, G.}, \bibinfo{year}{2023}.
\newblock \bibinfo{title}{Source inference attacks: Beyond membership inference attacks in federated learning}.
\newblock \bibinfo{journal}{IEEE Transactions on Dependable and Secure Computing} , \bibinfo{pages}{1--18}\DOIprefix\doi{10.1109/TDSC.2023.3321565}.
\bibitem[{Kairouz et~al.(2021)Kairouz, McMahan, Avent, Bellet, Bennis, Nitin~Bhagoji, Bonawitz, Charles, Cormode, Cummings, D’Oliveira, Eichner, El~Rouayheb, Evans, Gardner, Garrett, Gasc\'{o}n, Ghazi, Gibbons, Gruteser, Harchaoui, He, He, Huo, Hutchinson, Hsu, Jaggi, Javidi, Joshi, Khodak, Konecn\'{y}, Korolova, Koushanfar, Koyejo, Lepoint, Liu, Mittal, Mohri, Nock, \"{O}zg\"{u}r, Pagh, Qi, Ramage, Raskar, Raykova, Song, Song, Stich, Sun, Suresh, Tram\`{e}r, Vepakomma, Wang, Xiong, Xu, Yang, Yu, Yu and Zhao}]{Kairouz_2021_FL}
\bibinfo{author}{Kairouz, P.}, \bibinfo{author}{McMahan, H.B.}, \bibinfo{author}{Avent, B.}, \bibinfo{author}{Bellet, A.}, \bibinfo{author}{Bennis, M.}, \bibinfo{author}{Nitin~Bhagoji, A.}, \bibinfo{author}{Bonawitz, K.}, \bibinfo{author}{Charles, Z.}, \bibinfo{author}{Cormode, G.}, \bibinfo{author}{Cummings, R.}, \bibinfo{author}{D’Oliveira, R.G.L.}, \bibinfo{author}{Eichner, H.}, \bibinfo{author}{El~Rouayheb, S.}, \bibinfo{author}{Evans, D.}, \bibinfo{author}{Gardner, J.}, \bibinfo{author}{Garrett, Z.}, \bibinfo{author}{Gasc\'{o}n, A.}, \bibinfo{author}{Ghazi, B.}, \bibinfo{author}{Gibbons, P.B.}, \bibinfo{author}{Gruteser, M.}, \bibinfo{author}{Harchaoui, Z.}, \bibinfo{author}{He, C.}, \bibinfo{author}{He, L.}, \bibinfo{author}{Huo, Z.}, \bibinfo{author}{Hutchinson, B.}, \bibinfo{author}{Hsu, J.}, \bibinfo{author}{Jaggi, M.}, \bibinfo{author}{Javidi, T.}, \bibinfo{author}{Joshi, G.}, \bibinfo{author}{Khodak, M.}, \bibinfo{author}{Konecn\'{y}, J.}, \bibinfo{author}{Korolova, A.},
  \bibinfo{author}{Koushanfar, F.}, \bibinfo{author}{Koyejo, S.}, \bibinfo{author}{Lepoint, T.}, \bibinfo{author}{Liu, Y.}, \bibinfo{author}{Mittal, P.}, \bibinfo{author}{Mohri, M.}, \bibinfo{author}{Nock, R.}, \bibinfo{author}{\"{O}zg\"{u}r, A.}, \bibinfo{author}{Pagh, R.}, \bibinfo{author}{Qi, H.}, \bibinfo{author}{Ramage, D.}, \bibinfo{author}{Raskar, R.}, \bibinfo{author}{Raykova, M.}, \bibinfo{author}{Song, D.}, \bibinfo{author}{Song, W.}, \bibinfo{author}{Stich, S.U.}, \bibinfo{author}{Sun, Z.}, \bibinfo{author}{Suresh, A.T.}, \bibinfo{author}{Tram\`{e}r, F.}, \bibinfo{author}{Vepakomma, P.}, \bibinfo{author}{Wang, J.}, \bibinfo{author}{Xiong, L.}, \bibinfo{author}{Xu, Z.}, \bibinfo{author}{Yang, Q.}, \bibinfo{author}{Yu, F.X.}, \bibinfo{author}{Yu, H.}, \bibinfo{author}{Zhao, S.}, \bibinfo{year}{2021}.
\newblock \bibinfo{title}{Advances and open problems in federated learning}.
\newblock \bibinfo{journal}{Found. Trends Mach. Learn.} \bibinfo{volume}{14}, \bibinfo{pages}{1–210}.
\newblock \URLprefix \url{https://doi.org/10.1561/2200000083}, \DOIprefix\doi{10.1561/2200000083}.
\bibitem[{Kingma and Ba(2015)}]{Adam_Diederik_2015}
\bibinfo{author}{Kingma, D.P.}, \bibinfo{author}{Ba, J.}, \bibinfo{year}{2015}.
\newblock \bibinfo{title}{Adam: {A} method for stochastic optimization}, in: \bibinfo{booktitle}{Proceedings of the 3rd International Conference on Learning Representations}.
\bibitem[{LeCun et~al.(2006)LeCun, Chopra, Hadsell, Ranzato and Huang}]{lecun2006tutorial}
\bibinfo{author}{LeCun, Y.}, \bibinfo{author}{Chopra, S.}, \bibinfo{author}{Hadsell, R.}, \bibinfo{author}{Ranzato, M.}, \bibinfo{author}{Huang, F.}, \bibinfo{year}{2006}.
\newblock \bibinfo{title}{A tutorial on energy-based learning}.
\newblock \bibinfo{journal}{Predicting structured data} \bibinfo{volume}{1}.
\bibitem[{Li et~al.(2023)Li, Li and Ribeiro}]{li2023effective}
\bibinfo{author}{Li, J.}, \bibinfo{author}{Li, N.}, \bibinfo{author}{Ribeiro, B.}, \bibinfo{year}{2023}.
\newblock \bibinfo{title}{Effective passive membership inference attacks in federated learning against overparameterized models}, in: \bibinfo{booktitle}{Proceedings of the 11th International Conference on Learning Representations}.
\newblock \URLprefix \url{https://openreview.net/forum?id=QsCSLPP55Ku}.
\bibitem[{Liu and Nocedal(1989)}]{Dong_1989_LBFGS}
\bibinfo{author}{Liu, D.C.}, \bibinfo{author}{Nocedal, J.}, \bibinfo{year}{1989}.
\newblock \bibinfo{title}{On the limited memory {BFGS} method for large scale optimization}.
\newblock \bibinfo{journal}{Math. Program.} \bibinfo{volume}{45}, \bibinfo{pages}{503--528}.
\newblock \URLprefix \url{https://doi.org/10.1007/BF01589116}, \DOIprefix\doi{10.1007/BF01589116}.
\bibitem[{Liu et~al.(2023)Liu, Jiang and Zhu}]{Liu_23_SubjectMIA}
\bibinfo{author}{Liu, Y.}, \bibinfo{author}{Jiang, P.}, \bibinfo{author}{Zhu, L.}, \bibinfo{year}{2023}.
\newblock \bibinfo{title}{Subject-level membership inference attack via data augmentation and model discrepancy}.
\newblock \bibinfo{journal}{IEEE Transactions on Information Forensics and Security} \bibinfo{volume}{18}, \bibinfo{pages}{5848--5859}.
\newblock \DOIprefix\doi{10.1109/TIFS.2023.3318950}.
\bibitem[{Liu et~al.(2022)Liu, Zhao, Backes and Zhang}]{loss_trajectory_liu}
\bibinfo{author}{Liu, Y.}, \bibinfo{author}{Zhao, Z.}, \bibinfo{author}{Backes, M.}, \bibinfo{author}{Zhang, Y.}, \bibinfo{year}{2022}.
\newblock \bibinfo{title}{Membership inference attacks by exploiting loss trajectory}, in: \bibinfo{booktitle}{Proceedings of the 2022 ACM SIGSAC Conference on Computer and Communications Security}, \bibinfo{publisher}{Association for Computing Machinery}. p. \bibinfo{pages}{2085–2098}.
\newblock \URLprefix \url{https://doi.org/10.1145/3548606.3560684}, \DOIprefix\doi{10.1145/3548606.3560684}.
\bibitem[{Mahendran and Vedaldi(2015)}]{mahendran2015understanding}
\bibinfo{author}{Mahendran, A.}, \bibinfo{author}{Vedaldi, A.}, \bibinfo{year}{2015}.
\newblock \bibinfo{title}{Understanding deep image representations by inverting them}, in: \bibinfo{booktitle}{Proceedings of the IEEE conference on computer vision and pattern recognition}, \bibinfo{publisher}{IEEE Computer Society}. pp. \bibinfo{pages}{5188--5196}.
\bibitem[{Mahloujifar et~al.(2022)Mahloujifar, Ghosh and Chase}]{Mahloujifar_2022_PIA_from_poisoning}
\bibinfo{author}{Mahloujifar, S.}, \bibinfo{author}{Ghosh, E.}, \bibinfo{author}{Chase, M.}, \bibinfo{year}{2022}.
\newblock \bibinfo{title}{Property inference from poisoning}, in: \bibinfo{booktitle}{Proceedings of the 43th {IEEE} Symposium on Security and Privacy}, \bibinfo{publisher}{{IEEE} Computer Society}. pp. \bibinfo{pages}{1120--1137}.
\newblock \DOIprefix\doi{10.1109/SP46214.2022.9833623}.
\bibitem[{Marathe et~al.(2022)Marathe, Kanani and Peterson}]{marathe2022subject}
\bibinfo{author}{Marathe, V.}, \bibinfo{author}{Kanani, P.}, \bibinfo{author}{Peterson, D.W.}, \bibinfo{year}{2022}.
\newblock \bibinfo{title}{Subject level differential privacy with hierarchical gradient averaging}, in: \bibinfo{booktitle}{Workshop on Federated Learning: Recent Advances and New Challenges (in Conjunction with NeurIPS 2022)}.
\newblock \URLprefix \url{https://openreview.net/forum?id=vvQGXlHKkIG}.
\bibitem[{McMahan et~al.(2017)McMahan, Moore, Ramage, Hampson and y~Arcas}]{mcmahan2017communication}
\bibinfo{author}{McMahan, B.}, \bibinfo{author}{Moore, E.}, \bibinfo{author}{Ramage, D.}, \bibinfo{author}{Hampson, S.}, \bibinfo{author}{y~Arcas, B.A.}, \bibinfo{year}{2017}.
\newblock \bibinfo{title}{Communication-efficient learning of deep networks from decentralized data}, in: \bibinfo{booktitle}{Artificial intelligence and statistics}, \bibinfo{organization}{PMLR}. pp. \bibinfo{pages}{1273--1282}.
\bibitem[{McMahan et~al.(2018)McMahan, Ramage, Talwar and Zhang}]{brendan2018learning}
\bibinfo{author}{McMahan, H.B.}, \bibinfo{author}{Ramage, D.}, \bibinfo{author}{Talwar, K.}, \bibinfo{author}{Zhang, L.}, \bibinfo{year}{2018}.
\newblock \bibinfo{title}{Learning differentially private recurrent language models}, in: \bibinfo{booktitle}{Proceedings of the 6th International Conference on Learning Representations}.
\newblock \URLprefix \url{https://openreview.net/forum?id=BJ0hF1Z0b}.
\bibitem[{Melis et~al.(2019)Melis, Song, De~Cristofaro and Shmatikov}]{melis_2019_exploiting}
\bibinfo{author}{Melis, L.}, \bibinfo{author}{Song, C.}, \bibinfo{author}{De~Cristofaro, E.}, \bibinfo{author}{Shmatikov, V.}, \bibinfo{year}{2019}.
\newblock \bibinfo{title}{Exploiting unintended feature leakage in collaborative learning}, in: \bibinfo{booktitle}{Proceedings of the 40th {IEEE} Symposium on Security and Privacy}, \bibinfo{publisher}{{IEEE} Computer Society}. pp. \bibinfo{pages}{691--706}.
\bibitem[{Nasr et~al.(2019)Nasr, Shokri and Houmansadr}]{nasr_2019_comprehensive}
\bibinfo{author}{Nasr, M.}, \bibinfo{author}{Shokri, R.}, \bibinfo{author}{Houmansadr, A.}, \bibinfo{year}{2019}.
\newblock \bibinfo{title}{Comprehensive privacy analysis of deep learning: Passive and active white-box inference attacks against centralized and federated learning}, in: \bibinfo{booktitle}{Proceedings of the 40th {IEEE} Symposium on Security and Privacy}, \bibinfo{publisher}{{IEEE} Computer Society}. pp. \bibinfo{pages}{739--753}.
\bibitem[{Phong et~al.(2017)Phong, Aono, Hayashi, Wang and Moriai}]{Phong_2017_HE_on_FL}
\bibinfo{author}{Phong, L.T.}, \bibinfo{author}{Aono, Y.}, \bibinfo{author}{Hayashi, T.}, \bibinfo{author}{Wang, L.}, \bibinfo{author}{Moriai, S.}, \bibinfo{year}{2017}.
\newblock \bibinfo{title}{Privacy-preserving deep learning: Revisited and enhanced}, in: \bibinfo{booktitle}{Applications and Techniques in Information Security}, \bibinfo{publisher}{Springer Singapore}. pp. \bibinfo{pages}{100--110}.
\bibitem[{Pichler et~al.(2022)Pichler, Romanelli, Vega and Piantanida}]{pichler2022perfectly}
\bibinfo{author}{Pichler, G.}, \bibinfo{author}{Romanelli, M.}, \bibinfo{author}{Vega, L.R.}, \bibinfo{author}{Piantanida, P.}, \bibinfo{year}{2022}.
\newblock \bibinfo{title}{Perfectly accurate membership inference by a dishonest central server in federated learning}.
\newblock \bibinfo{journal}{arXiv preprint arXiv:2203.16463} .
\bibitem[{Platforms(2024)}]{opacus_2024}
\bibinfo{author}{Platforms, M.}, \bibinfo{year}{2024}.
\newblock \bibinfo{title}{Opacus}.
\newblock \bibinfo{howpublished}{\url{https://opacus.ai/}}.
\newblock \bibinfo{note}{Accessed: 2024-05-27}.
\bibitem[{Qi et~al.(2024)Qi, Chiaro, Guzzo, Ianni, Fortino and Piccialli}]{QI_2024_FLaberage}
\bibinfo{author}{Qi, P.}, \bibinfo{author}{Chiaro, D.}, \bibinfo{author}{Guzzo, A.}, \bibinfo{author}{Ianni, M.}, \bibinfo{author}{Fortino, G.}, \bibinfo{author}{Piccialli, F.}, \bibinfo{year}{2024}.
\newblock \bibinfo{title}{Model aggregation techniques in federated learning: A comprehensive survey}.
\newblock \bibinfo{journal}{Future Generation Computer Systems} \bibinfo{volume}{150}, \bibinfo{pages}{272--293}.
\newblock \URLprefix \url{https://www.sciencedirect.com/science/article/pii/S0167739X23003333}, \DOIprefix\doi{https://doi.org/10.1016/j.future.2023.09.008}.
\bibitem[{Shokri et~al.(2017)Shokri, Stronati, Song and Shmatikov}]{shokri_membership_2017}
\bibinfo{author}{Shokri, R.}, \bibinfo{author}{Stronati, M.}, \bibinfo{author}{Song, C.}, \bibinfo{author}{Shmatikov, V.}, \bibinfo{year}{2017}.
\newblock \bibinfo{title}{Membership inference attacks against machine learning models}, in: \bibinfo{booktitle}{Proceedings of the 38th {IEEE} Symposium on Security and Privacy}, \bibinfo{publisher}{{IEEE} Computer Society}. pp. \bibinfo{pages}{3--18}.
\newblock \URLprefix \url{https://doi.org/10.1109/SP.2017.41}, \DOIprefix\doi{10.1109/SP.2017.41}.
\bibitem[{Song and Shmatikov(2020)}]{Song2020Overlearning}
\bibinfo{author}{Song, C.}, \bibinfo{author}{Shmatikov, V.}, \bibinfo{year}{2020}.
\newblock \bibinfo{title}{Overlearning reveals sensitive attributes}, in: \bibinfo{booktitle}{Proceedings of the 8th International Conference on Learning Representations}.
\newblock \URLprefix \url{https://openreview.net/forum?id=SJeNz04tDS}.
\bibitem[{Song and Mittal(2021)}]{song_systematic_2021}
\bibinfo{author}{Song, L.}, \bibinfo{author}{Mittal, P.}, \bibinfo{year}{2021}.
\newblock \bibinfo{title}{Systematic evaluation of privacy risks of machine learning models}, in: \bibinfo{booktitle}{Proceedings of the 30th {USENIX} Security Symposium}, \bibinfo{publisher}{{USENIX} Association}. pp. \bibinfo{pages}{2615--2632}.
\bibitem[{Suri and Evans(2022)}]{suri2022formalizing}
\bibinfo{author}{Suri, A.}, \bibinfo{author}{Evans, D.}, \bibinfo{year}{2022}.
\newblock \bibinfo{title}{Formalizing and estimating distribution inference risks}.
\newblock \bibinfo{journal}{Proceedings on Privacy Enhancing Technologies} \bibinfo{volume}{2022}, \bibinfo{pages}{528--551}.
\newblock \DOIprefix\doi{10.56553/POPETS-2022-0121}.
\bibitem[{Suri et~al.(2022)Suri, Kanani, Marathe and Peterson}]{suri2022subject}
\bibinfo{author}{Suri, A.}, \bibinfo{author}{Kanani, P.}, \bibinfo{author}{Marathe, V.J.}, \bibinfo{author}{Peterson, D.W.}, \bibinfo{year}{2022}.
\newblock \bibinfo{title}{Subject membership inference attacks in federated learning}.
\newblock \bibinfo{journal}{arXiv preprint arXiv:2206.03317} .
\bibitem[{Suri et~al.(2023)Suri, Lu, Chen and Evans}]{Suri_2023_DIA}
\bibinfo{author}{Suri, A.}, \bibinfo{author}{Lu, Y.}, \bibinfo{author}{Chen, Y.}, \bibinfo{author}{Evans, D.}, \bibinfo{year}{2023}.
\newblock \bibinfo{title}{Dissecting distribution inference}, in: \bibinfo{booktitle}{2023 IEEE Conference on Secure and Trustworthy Machine Learning (SaTML)}, \bibinfo{publisher}{IEEE Computer Society}. pp. \bibinfo{pages}{150--164}.
\newblock \URLprefix \url{https://doi.ieeecomputersociety.org/10.1109/SaTML54575.2023.00019}, \DOIprefix\doi{10.1109/SaTML54575.2023.00019}.
\bibitem[{Szegedy et~al.(2014)Szegedy, Zaremba, Sutskever, Bruna, Erhan, Goodfellow and Fergus}]{Christian_properties_2014}
\bibinfo{author}{Szegedy, C.}, \bibinfo{author}{Zaremba, W.}, \bibinfo{author}{Sutskever, I.}, \bibinfo{author}{Bruna, J.}, \bibinfo{author}{Erhan, D.}, \bibinfo{author}{Goodfellow, I.J.}, \bibinfo{author}{Fergus, R.}, \bibinfo{year}{2014}.
\newblock \bibinfo{title}{Intriguing properties of neural networks}, in: \bibinfo{booktitle}{Proceedings of the 2nd International Conference on Learning Representations}.
\bibitem[{TEAM(2017)}]{GDPR_citation}
\bibinfo{author}{TEAM, I.G.P.}, \bibinfo{year}{2017}.
\newblock \bibinfo{title}{EU General Data Protection Regulation (GDPR): An Implementation and Compliance Guide - Second edition}.
\newblock \bibinfo{edition}{2} ed., \bibinfo{publisher}{IT Governance Publishing}.
\newblock \URLprefix \url{http://www.jstor.org/stable/j.ctt1trkk7x}.
\bibitem[{Wang and Gong(2018)}]{Wang_2018_hypersteal}
\bibinfo{author}{Wang, B.}, \bibinfo{author}{Gong, N.}, \bibinfo{year}{2018}.
\newblock \bibinfo{title}{Stealing hyperparameters in machine learning}, in: \bibinfo{booktitle}{Proceedings of the 39th {IEEE} Symposium on Security and Privacy}, \bibinfo{publisher}{IEEE Computer Society}. pp. \bibinfo{pages}{36--52}.
\newblock \URLprefix \url{https://doi.ieeecomputersociety.org/10.1109/SP.2018.00038}, \DOIprefix\doi{10.1109/SP.2018.00038}.
\bibitem[{Wang et~al.(2023)Wang, Huang, Song, Wu, Xue and Ren}]{Wang_2023_PAPIA}
\bibinfo{author}{Wang, Z.}, \bibinfo{author}{Huang, Y.}, \bibinfo{author}{Song, M.}, \bibinfo{author}{Wu, L.}, \bibinfo{author}{Xue, F.}, \bibinfo{author}{Ren, K.}, \bibinfo{year}{2023}.
\newblock \bibinfo{title}{Poisoning-assisted property inference attack against federated learning}.
\newblock \bibinfo{journal}{IEEE Transactions on Dependable and Secure Computing} \bibinfo{volume}{20}, \bibinfo{pages}{3328--3340}.
\newblock \DOIprefix\doi{10.1109/TDSC.2022.3196646}.
\bibitem[{Wang et~al.(2019)Wang, Song, Zhang, Song, Wang and Qi}]{Wang_2019_User_level}
\bibinfo{author}{Wang, Z.}, \bibinfo{author}{Song, M.}, \bibinfo{author}{Zhang, Z.}, \bibinfo{author}{Song, Y.}, \bibinfo{author}{Wang, Q.}, \bibinfo{author}{Qi, H.}, \bibinfo{year}{2019}.
\newblock \bibinfo{title}{Beyond inferring class representatives: User-level privacy leakage from federated learning}, in: \bibinfo{booktitle}{IEEE INFOCOM 2019 - IEEE Conference on Computer Communications}, \bibinfo{publisher}{IEEE Press}. p. \bibinfo{pages}{2512–2520}.
\newblock \URLprefix \url{https://doi.org/10.1109/INFOCOM.2019.8737416}, \DOIprefix\doi{10.1109/INFOCOM.2019.8737416}.
\bibitem[{Wattenberg et~al.(2016)Wattenberg, Viégas and Johnson}]{wattenberg2016how}
\bibinfo{author}{Wattenberg, M.}, \bibinfo{author}{Viégas, F.}, \bibinfo{author}{Johnson, I.}, \bibinfo{year}{2016}.
\newblock \bibinfo{title}{How to use t-sne effectively}.
\newblock \bibinfo{journal}{Distill} \URLprefix \url{http://distill.pub/2016/misread-tsne}, \DOIprefix\doi{10.23915/distill.00002}.
\bibitem[{Zari et~al.(2021)Zari, Xu and Neglia}]{zari_2021_efficient}
\bibinfo{author}{Zari, O.}, \bibinfo{author}{Xu, C.}, \bibinfo{author}{Neglia, G.}, \bibinfo{year}{2021}.
\newblock \bibinfo{title}{Efficient passive membership inference attack in federated learning}.
\newblock \bibinfo{journal}{arXiv preprint arXiv:2111.00430} .
\bibitem[{Zhang et~al.(2020)Zhang, Zhang, Chen and Yu}]{Zhang_2022_GANMIA}
\bibinfo{author}{Zhang, J.}, \bibinfo{author}{Zhang, J.}, \bibinfo{author}{Chen, J.}, \bibinfo{author}{Yu, S.}, \bibinfo{year}{2020}.
\newblock \bibinfo{title}{Gan enhanced membership inference: A passive local attack in federated learning}, in: \bibinfo{booktitle}{ICC 2020 - 2020 IEEE International Conference on Communications (ICC)}, \bibinfo{publisher}{IEEE Computer Society}. pp. \bibinfo{pages}{1--6}.
\newblock \DOIprefix\doi{10.1109/ICC40277.2020.9148790}.
\bibitem[{Zhang et~al.(2023)Zhang, Li, Li, Cai, Gao, Dou and Chen}]{Zhang_2023_efficient_MIA}
\bibinfo{author}{Zhang, L.}, \bibinfo{author}{Li, L.}, \bibinfo{author}{Li, X.}, \bibinfo{author}{Cai, B.}, \bibinfo{author}{Gao, Y.}, \bibinfo{author}{Dou, R.}, \bibinfo{author}{Chen, L.}, \bibinfo{year}{2023}.
\newblock \bibinfo{title}{Efficient membership inference attacks against federated learning via bias differences}, in: \bibinfo{booktitle}{Proceedings of the 26th International Symposium on Research in Attacks, Intrusions and Defenses}, \bibinfo{publisher}{Association for Computing Machinery}. p. \bibinfo{pages}{222–235}.
\newblock \URLprefix \url{https://doi.org/10.1145/3607199.3607204}, \DOIprefix\doi{10.1145/3607199.3607204}.
\bibitem[{Zhang et~al.(2021)Zhang, Tople and Ohrimenko}]{zhang2021leakage}
\bibinfo{author}{Zhang, W.}, \bibinfo{author}{Tople, S.}, \bibinfo{author}{Ohrimenko, O.}, \bibinfo{year}{2021}.
\newblock \bibinfo{title}{Leakage of dataset properties in $\{$Multi-Party$\}$ machine learning}, in: \bibinfo{booktitle}{Proceedings of the 30th {USENIX} Security Symposium}, \bibinfo{publisher}{{USENIX} Association}. pp. \bibinfo{pages}{2687--2704}.
\bibitem[{Zhao et~al.(2020)Zhao, Mopuri and Bilen}]{zhao_2020_idlg}
\bibinfo{author}{Zhao, B.}, \bibinfo{author}{Mopuri, K.R.}, \bibinfo{author}{Bilen, H.}, \bibinfo{year}{2020}.
\newblock \bibinfo{title}{idlg: Improved deep leakage from gradients}.
\newblock \bibinfo{journal}{arXiv preprint arXiv:2001.02610} .
\bibitem[{Zhao et~al.(2021)Zhao, Chen, Zhang, Yang, Tu, Han, Zhu and Chen}]{Zhao_user-level_2021}
\bibinfo{author}{Zhao, Y.}, \bibinfo{author}{Chen, J.}, \bibinfo{author}{Zhang, J.}, \bibinfo{author}{Yang, Z.}, \bibinfo{author}{Tu, H.}, \bibinfo{author}{Han, H.}, \bibinfo{author}{Zhu, K.}, \bibinfo{author}{Chen, B.}, \bibinfo{year}{2021}.
\newblock \bibinfo{title}{User-level membership inference for federated learning in wireless network environment}.
\newblock \bibinfo{journal}{Wireless Communications and Mobile Computing} \bibinfo{volume}{2021}, \bibinfo{pages}{5534270}.
\newblock \URLprefix \url{https://doi.org/10.1155/2021/5534270}, \DOIprefix\doi{10.1155/2021/5534270}.
\bibitem[{Zhu et~al.(2019)Zhu, Liu and Han}]{Zhu_2019_Deep_leakage}
\bibinfo{author}{Zhu, L.}, \bibinfo{author}{Liu, Z.}, \bibinfo{author}{Han, S.}, \bibinfo{year}{2019}.
\newblock \bibinfo{title}{Deep leakage from gradients}, in: \bibinfo{booktitle}{Advances in Neural Information Processing Systems 32}, pp. \bibinfo{pages}{14747--14756}.

\end{thebibliography}





\end{document}